\documentclass[aps,prd,floatfix,preprintnumbers]{revtex4}

\usepackage[english]{babel}
\usepackage{amsmath,amssymb}
\usepackage[dvips]{graphicx}
\usepackage{rotating}

\def\rpv{$R_p \hspace{-1em}/\;\:\hspace{0.2em}$}

\def\lsim{\raise0.3ex\hbox{$\;<$\kern-0.75em\raise-1.1ex\hbox{$\sim\;$}}}
\def\vb#1{\vbox to #1 pt{}}
\def\eslash{$E \hspace{-0.9em}/\;\:$ }

\newcommand{\AddrAHEP}{%
  AHEP Group, Institut de F\'{\i}sica Corpuscular --
  C.S.I.C. \& Universitat de Val{\`e}ncia \\
  Edificio Institutos de Paterna, Apt 22085, E--46071 Valencia, Spain}
\newcommand{\AddrWur}{%
Institut f\"ur Theoretische Physik und Astronomie, 
Universit\"at W\"urzburg\\
Am Hubland, 
97074 Wuerzburg}

\begin{document}

\preprint{IFIC/08-12}
\title{Spontaneous R-parity violation: Lightest neutralino decays 
and neutrino mixing angles at future colliders}

\author{M.~Hirsch} \email{mahirsch@ific.uv.es} \affiliation{\AddrAHEP}

\author{W.\ Porod} \email{porod@physik.uni-wuerzburg.de}
\affiliation{\AddrWur}

\author{A.~Vicente} \email{Avelino.Vicente@ific.uv.es} \affiliation{\AddrAHEP}


\begin{abstract}
\vspace*{1cm}

We study the decays of the lightest supersymmetric particle (LSP) 
in models with spontaneously broken R-parity. We focus on the two 
cases that the LSP is either a bino or a neutral singlet lepton. 
We work out the most important phenomenological differences between 
these two scenarios and discuss also how they might be distinguished 
from explicit R-Parity breaking models. In both cases we find that 
certain ratios of decay branching ratios are correlated with either 
the solar or the atmospheric (and reactor) neutrino angle. The 
hypothesis that spontaneous R-Parity violation is the source of 
the observed neutrino masses is therefore potentially testable 
at the LHC.

\end{abstract}

\maketitle

\section{Introduction}

Experiments on solar and atmospheric neutrinos have demonstrated 
that neutrinos have non-zero masses and mixing angles  
\cite{Fukuda:1998mi,Ahmad:2002jz,Eguchi:2002dm}. Indeed, with the 
latest experimental data by the MINOS \cite{Collaboration:2007zza} 
and KamLAND \cite{KamLAND2007} collaborations neutrino oscillation 
experiments truly have entered the precision era \cite{Maltoni:2004ei}. 

Many models have been proposed, which potentially can explain these data. 
Certainly the most popular are variants based on the see-saw mechanism 
\cite{Minkowski:1977sc,Yanagida:1979as,Gell-Mann:1980vs,Mohapatra:1979ia%
,Schechter:1980gr}. However, the classical seesaw mechanism puts the 
scale of lepton number violation near the grand unification scale and, 
therefore, can not be directly tested. Indirect tests of the seesaw 
might be possible, if low-scale supersymmetry is realized in nature 
\cite{Deppisch:2002vz,Blair:2002pg,Freitas:2005et,Buckley:2006nv%
,Deppisch:2007xu}. 
Measurements of 
masses and branching ratios of supersymmetric particles can be used to 
obtain indirect information on the range of allowed seesaw parameters, 
if some specific scenarios for the soft-breaking parameters are assumed 
\cite{Blair:2002pg,Freitas:2005et,Buckley:2006nv}. 
On the other hand, a number of neutrino mass models exist, in which 
neutrino masses are generated at the electro-weak scale. Examples 
include the Babu-Zee model \cite{Babu:1988ki}, Leptoquarks 
\cite{AristizabalSierra:2007nf} or in case of supersymmetry the 
breaking of R-parity \cite{Hall:1984id}. These low energy models 
usually lead to testable predictions for future colliders. For example, 
in the case of R-parity violation (\rpv), in particular for  models based 
on bilinear R-parity breaking terms \cite{Hirsch:2004he}, correlations 
between certain branching ratios of the lightest supersymmetric particle 
decays and the measured neutrino mixing angles have been found 
\cite{Porod:2000hv,Restrepo:2001me,Hirsch:2002ys,Bartl:2003uq,Hirsch:2003fe}. 

In \rpv the {\em neutralino} as a dark matter candidate is lost. Recent 
WMAP data \cite{Spergel:2006hy}, however, have confirmed the existence 
of non-baryonic dark matter and measured its contribution to the energy 
budget of the universe with unprecedented accuracy. Thus, in \rpv 
one needs a non-standard explanation of dark matter (DM). Examples for DM candidates 
in \rpv include (i) light gravitinos 
\cite{Borgani:1996ag,Takayama:2000uz,Hirsch:2005ag},
(ii) the axion \cite{Kim:1986ax,Raffelt:1996wa} or (iii) its superpartner, 
the axino \cite{Chun:1999cq,Hirsch:2005jn}, to mention a few.

Whether R-parity is conserved or not can, in principle, be easily 
decided in the case of explicit R-parity violation since (a) neutrino 
physics implies that the lightest supersymmetric particle (LSP) will 
decay inside a typical detector of existing and future high energy
experiments \cite{Porod:2000hv,Hirsch:2003fe} and (b) the branching 
fraction for (completely) invisible LSP decays is at most ${\cal O}(10 \%)$ 
and typically smaller\cite{Porod:2000hv,Hirsch:2006di}. {\em Spontaneous} 
violation of R-parity (s-\rpv)
\cite{Aulakh:1982yn,Masiero:1990uj}, on the other hand, implies the 
existence of a Goldstone boson, the majoron (J). In s-\rpv 
the lightest neutralino can then decay according to $\chi^0
\rightarrow J + \nu$, i.e.~completely invisible. It has been shown 
\cite{Hirsch:2006di} that this decay mode can in fact be the dominant 
one, with branching ratios close to 100 \%, despite the smallness of 
neutrino masses, in case the scale of R-parity breaking is relatively 
low. In this limit, the accelerator phenomenology of models with 
spontaneous violation of R-parity can resemble the MSSM with conserved 
R-parity and large statistics might be necessary before it can be 
established that R-parity indeed is broken. 

In this paper we study scenarios where the lightest neutralino is the 
LSP focusing on two representative examples: (i) a bino-like LSP as 
this case has been extensively studied in the literature, both in case 
of conserved R-parity as well as (explicitly) broken R-parity. We will 
re-iterate our previous result \cite{Hirsch:2006di} that there are regions 
in parameter space, where it will be difficult to obtain clarity about 
the underlying model in the first years of LHC. We will also discuss, 
how measurements of branching ratios can lead to tests of the model 
as the origin of the neutrino mass, in case sufficient statistics 
for the final states with charged leptons can be obtained. (ii) A 
singlino-like LSP. This case has not been studied in the past but 
it is the only part of the allowed parameter space where singlinos 
can be produced and studied at future accelerators. In addition, this 
case allows interesting cross-checks with respect to neutrino physics 
different from a bino LSP.

The paper is organized as follows: in Sect.~\ref{sec:model} we present 
the model with special emphasis on neutrino physics as well as the most 
important couplings of the lightest neutralino. In Sect.~\ref{sec:pheno} 
we discuss production and decays of the lightest neutralino at the LHC 
as well as a future International Linear Collider (ILC) putting some 
emphasis on how to distinguish bino and singlino LSPs. In 
Sect.~\ref{sec:corr} we discuss the different correlations between 
ratios of neutralino decay branching ratios and neutrino mixing angles. 
Finally, we draw our conclusions in Sect.~\ref{sec:con}.

\section{Spontaneous R-parity violation}
\label{sec:model}

\subsection{Model basics}
\label{sec:mbasic}

Spontaneous breaking of a global symmetry leads to a Goldstone boson,  
in case of lepton number breaking usually called the Majoron. Spontaneous 
breaking of R-parity through left sneutrinos \cite{Aulakh:1982yn}, 
produces a doublet Majoron, which is ruled out by LEP and astrophysical 
data \cite{Raffelt:1996wa,Yao:2006px}. To construct a phenomenologically 
consistent version of s-\rpv it is therefore necessary to extend the 
particle content of the MSSM by at least one singlet field, $\widehat\nu^c$, 
which carries lepton number. For reasons to be explained in more 
detail below, the model we consider \cite{Masiero:1990uj} contains three 
additional singlet superfields, namely, $\widehat\nu^c$, $\widehat S$ and 
$\widehat\Phi$, with lepton number assignments of $L=-1,1,0$ 
respectively. 

The superpotential can be written as \cite{Masiero:1990uj}
\begin{eqnarray} %
{\cal W} &=& h_U^{ij}\widehat Q_i \widehat U_j\widehat H_u
          +  h_D^{ij}\widehat Q_i\widehat D_j\widehat H_d
          +  h_E^{ij}\widehat L_i\widehat E_j\widehat H_d \nonumber
\\
        & + & h_{\nu}^{i}\widehat L_i\widehat \nu^c\widehat H_u
          - h_0 \widehat H_d \widehat H_u \widehat\Phi
          + h \widehat\Phi \widehat\nu^c\widehat S +
          \frac{\lambda}{3!} \widehat\Phi^3 .
\label{eq:Wsuppot}
\end{eqnarray}
The basic guiding principle in the construction of Eq.~(\ref{eq:Wsuppot})
is that lepton number is conserved at the level of the superpotential.
The first three terms are the usual MSSM Yukawa terms. The terms
coupling the lepton doublets to $\widehat\nu^c$ fix lepton number.
The coupling of the field $\widehat\Phi$ with the Higgs doublets
generates an effective $\mu$-term a l\'a Next to Minimal Supersymmetric
Standard Model \cite{Barbieri:1982eh}. The last two terms, involving
only singlet fields, give mass to $\widehat\nu^c$, $\widehat S$ and
$\widehat\Phi$, once $\Phi$ develops a vacuum expectation value (vev).

For simplicity we consider only one generation of $\widehat\nu^c$ and
$\widehat S$. Adding more generations of $\widehat\nu^c$ or $\widehat S$
does not add any qualitatively new features to the model. Note also,
that the superpotential, Eq.~(\ref{eq:Wsuppot}), does not contain any
terms with dimension of mass, thus potentially offering a solution
to the $\mu$-problem of supersymmetry. The inclusion of $\widehat S$ allows 
to generate a ``Dirac''-like mass term for $\widehat\nu^c$, once 
$\widehat\Phi$ gets a vev. The soft supersymmetry breaking
terms of this model can be found in \cite{Hirsch:2004rw}.

At low energy, i.e.~after electro-weak symmetry breaking, various
fields acquire vevs. Besides the usual MSSM Higgs boson vevs $v_d$ and
$v_u$, these are $\langle \Phi \rangle = v_{\Phi}/\sqrt{2}$,
$\langle {\tilde \nu}^c \rangle = v_R/\sqrt{2}$,
$\langle {\tilde S} \rangle = v_S/\sqrt{2}$ and
$\langle {\tilde \nu}_i \rangle = v_{L_i}/\sqrt{2}$.
Note, that $v_R \ne 0$ generates effective bilinear terms
$\epsilon_i = h_{\nu}^i v_R/\sqrt{2}$ and that $v_R$, $v_S$ and $v_{L_i}$
violate lepton number as well as R-parity.

\subsection{Neutralino-neutrino mass matrix}
\label{sec:neutralino}

\noindent
In the basis 
\begin{equation}
(-i\lambda',-i\lambda^3,{\tilde H_d},{\tilde H_u},\nu_e,\nu_{\mu},\nu_{\tau},
\nu^c,S,\tilde{\Phi})
\label{eq:defbasis}
\end{equation}
the mass matrix of the neutral fermions following from Eq. (\ref{eq:Wsuppot}) 
can be written as \cite{Hirsch:2004rw,Hirsch:2005wd} 

\begin{equation}
\mathbf{M_N}=
\left(\begin{array}{lllll}
\mathbf{M_{\chi^0}} & \mathbf{m_{\chi^0\nu}}& \mathbf{m_{\chi^0\nu^c}}& 
\mathbf{0}& \mathbf{m_{\chi^0\Phi} } \\
\\
\mathbf{m^T_{\chi^0\nu}} & \mathbf{0} & \mathbf{m_{D}} & 
\mathbf{0} & \mathbf{0} \\
\\
\mathbf{m^T_{\chi^0\nu^c}}&\mathbf{m^T_{D}} & \mathbf{0} &
\mathbf{M_{\nu^c S}} & \mathbf{M_{\nu^c\Phi}} \\
\\
\mathbf{0} &\mathbf{0} &
\mathbf{M_{\nu^c S}} &\mathbf{0} &\mathbf{M_{S\Phi}} \\
\\
\mathbf{m^T_{\chi^0\Phi} } & \mathbf{0} & \mathbf{M_{\nu^c\Phi}} & 
\mathbf{M_{S\Phi}} & \mathbf{M_{\Phi}}
\end{array} \right).
\label{eq:mass}
\end{equation}
\noindent
Eq. (\ref{eq:mass}) can be diagonalized in the standard way,
\begin{equation}
\mathbf{\widehat M_N}= {\cal N}^* \mathbf{M_N}{\cal N}^{-1}.
\label{eq:diag mass}
\end{equation}
We have chosen the basis in eq. (\ref{eq:defbasis}), such that 
${\cal N}$ reduces to the MSSM neutralino rotation matrix in 
the limit where (a) R-parity is conserved and (b) the field $\Phi$ 
is decoupled.
The various sub-blocks in eq. (\ref{eq:mass}) are defined as 
follows. The matrix $\mathbf{M_{\chi^0}}$ is the standard MSSM 
neutralino mass matrix:
\begin{equation}
\mathbf{M_{\chi^0}} =
\left(\begin{array}{llll}
M_1 & 0    & - \frac{1}{2} g' v_d & + \frac{1}{2} g' v_u \\ \vb{12}
0   & M_ 2 & + \frac{1}{2} g v_d &  - \frac{1}{2} g v_u \\ \vb{12}
- \frac{1}{2} g' v_d & + \frac{1}{2} g v_d &    0 & -\mu \\ \vb{12}
+ \frac{1}{2} g' v_u & - \frac{1}{2} g v_u & -\mu & 0
\end{array} \right).
\label{eq:mntrl}
\end{equation}
Here, $\mu =h_0v_{\Phi}/\sqrt{2}$. $\mathbf{m_{\chi^0\nu}}$ is the 
R-parity violating neutrino-neutralino mixing part, which also appears 
in explicit bilinear R-parity breaking models:
\begin{equation}
\mathbf{m^T_{\chi^0\nu}} =
\left(\begin{array}{llll}
-\frac{1}{2}g'v_{L e} &\frac{1}{2}gv_{L e} & 0 & \epsilon_e \\[2mm]
-\frac{1}{2}g'v_{L \mu}& \frac{1}{2}gv_{L \mu}& 0& \epsilon_{\mu}\\[2mm]
-\frac{1}{2}g'v_{L \tau} & \frac{1}{2}gv_{L \tau} & 0& \epsilon_{\tau} 
\end{array} \right),
\label{eq:mrpv}
\end{equation}
where $v_{L i}$ are the vevs of the left-sneutrinos.

\noindent
$\mathbf{m_{\chi^0\nu^c}}$ is given as
\begin{equation}
\mathbf{m^T_{\chi^0\nu^c}} =
\left(
0, \hskip2mm 0, \hskip2mm  0, \hskip2mm 
\frac{1}{\sqrt{2}}\sum h_{\nu}^{i} v_{L i} 
\right).
\label{eq:mchinuc}
\end{equation}
and $\mathbf{m^T_{\chi^0\Phi} }$ is 
\begin{equation}
\mathbf{m^T_{\chi^0\Phi} }
= ( 0 , 0, - \frac{1}{\sqrt{2}}h_0 v_u , - \frac{1}{\sqrt{2}}h_0 v_d).
\label{eq:mchiphi}
\end{equation}
The ``Dirac'' mass matrix is defined in the usual way:
\begin{equation}
(\mathbf{m_{D}})_{i} = \frac{1}{\sqrt{2}}h_{\nu}^{i}v_u.
\label{eq:mD}
\end{equation}
And, finally, 
\begin{equation}
(\mathbf{M_{\nu^c S}}) = \frac{1}{\sqrt{2}}h v_{\Phi}, \hskip5mm 
\mathbf{M_{\nu^c\Phi}} = \frac{1}{\sqrt{2}} h v_S, \hskip5mm 
\mathbf{M_{S\Phi}} = \frac{1}{\sqrt{2}} h v_R, \hskip5mm 
\mathbf{M_{\Phi}} = \frac{\lambda}{\sqrt{2}} v_{\Phi}.
\label{eq:mp_other}
\end{equation}
The matrix, eq. (\ref{eq:mass}), produces ten eigenvalues with 
vastly different masses. First, since $Det(\mathbf{M_N})=0$, one 
state is massless at tree-level. Then there are two more very light 
states, together they form to a good approximation the three observed, 
light doublet neutrinos. Their masses and mixing will be discussed in 
detail in the next subsection. 

The remaining seven eigenstates are typically heavy. They can be 
sub-divided into two groups: Mainly doublet and mainly singlet states. 
There are usually four states which are very similar to the well-known 
MSSM neutralinos. Unless $h_0$ is large {\em and} $\lambda v_{\Phi}$ 
small, mixing between the phino and the higgsinos is small  
\footnote{As in the NMSSM, if the field $\Phi$ is light and the coupling 
$h_0$ large, one has five neutralino states}, and there are three 
singlets. From these singlets, unless $(h-\lambda)\le v_R/v_{\Phi}$, 
$\nu^c$ and $S$ form a quasi-Dirac pair, which we will loosely call 
``the singlino'', 
${\cal S}_{1,2} \simeq \frac{1}{\sqrt{2}} (\nu^c \mp S)$. Note, that
this is a different state compared to the NMSSM singlino \cite{Franke:2001nx}
which corresponds to $\tilde \Phi$ in our notation.

Which of the seven, heavy states is the lightest depends on a number 
of unknown parameters and can not be predicted. In our analysis below 
we will concentrate on two cases: (a) As in mSugra motivated 
scenarios $M_1$ is the smallest parameter and the lightest state 
mainly a bino. We study this case in order to work out the 
differences to (i) the well-studied phenomenology of the MSSM; 
and (ii) to the explicit R-parity violating case studied in 
\cite {Porod:2000hv}. The second case we consider is (b) the 
singlino $\cal S$ being the lightest state. This case is 
interesting, since it is the only part of the parameter space, 
where singlets indeed can be produced and studied at accelerators.

\subsection{Neutrino masses}
\label{sec:neutrino-masses}

Since neutrino masses are much smaller than all other fermion mass 
terms, one can find the effective neutrino mass matrix in a 
seesaw--type approximation \cite{Hirsch:2004rw,Hirsch:2005wd}. 
First we define the small expansion parameters $\xi_{ij}$, which 
characterize the mixing between the neutrino sector and the 
seven heavy neutral fermion states, the ``neutralinos'' of the 
model, 
\begin{equation}\label{eq:defxi}
\xi=\mathbf{m}_{3\times 7}\cdot \mathbf{M_H}^{-1}.
\end{equation}
The sub-matrix describing the seven heavy states of eq. (\ref{eq:mass}) 
is
\begin{equation}
\mathbf{M_H}=
\left(\begin{array}{llll}
\mathbf{M_{\chi^0}} & \mathbf{0} & 
\mathbf{0}& \mathbf{m_{\chi^0\Phi} } \\
\\
\mathbf{0}& \mathbf{0} &
\mathbf{M_{\nu^c S}} & \mathbf{M_{\nu^c\Phi}} \\
\\
\mathbf{0} &
\mathbf{M_{\nu^c S}} &\mathbf{0} &\mathbf{M_{S\Phi}} \\
\\
\mathbf{m^T_{\chi^0\Phi} } & \mathbf{M_{\nu^c\Phi}} & 
\mathbf{M_{S\Phi}} & \mathbf{M_{\Phi}}
\end{array} \right).
\label{eq:mass77}
\end{equation}
\noindent
and
\begin{equation}
\mathbf{m}_{3\times 7}=
\left(\begin{array}{llll}
\mathbf{m^T_{\chi^0\nu}} & \mathbf{m_{D}} & 
\mathbf{0} & \mathbf{0} 
\end{array} \right).
\label{eq:mass37}
\end{equation}
\noindent
We have neglected $\mathbf{m_{\chi^0\nu^c}}$ in eq. (\ref {eq:mass77}) 
since it is doubly suppressed. The ``effective'' ($3,3$) neutrino mass 
matrix is then given in seesaw approximation by
\begin{equation}\label{eq:defmnueff}
(\boldsymbol{m_{\nu\nu}^{\rm eff}}) =
- \mathbf{m}_{3\times 7}\cdot \mathbf{M_H}^{-1}\mathbf{m}_{3\times 7}^T.
\end{equation}
In the following we will use the symbol $N_{ij}$ with $i,j=1,...7$ as 
the matrix which diagonalizes eq. (\ref{eq:mass77}). Our $N$ reduces 
to the MSSM neutralino mixing matrix $N$, in the limit where the singlets 
decouple, i.e. $h_0\to 0$ or $M_{\Phi}\to \infty$. After some straightforward 
algebra $\xi_{ij}$ can be written as 
\begin{equation}\label{def:xi}
\xi_{ij}=K_\Lambda^j \Lambda_i + K_\epsilon^j \epsilon_i,
\end{equation}
where the effective bilinear R--parity violating parameters 
$\epsilon_{i}$ and $\Lambda_i$ are 
\begin{equation}
  \label{eq:eps}
\epsilon_{i} = h_{\nu}^{i}\, \frac{v_R}{\sqrt{2}}  
\end{equation}
and
\begin{equation}
\Lambda_i = \epsilon_i v_d + \mu v_{L_i}. 
\label{eq:deflam0}
\end{equation}
The coefficients $K$ are given as 
\begin{eqnarray}\label{defK}
K_\Lambda^1 = -\frac{2 g' M_2 \mu}{m_\gamma}a, 
&\hskip5mm & K_\epsilon^1 = -\frac{2 g' M_2 \mu}{m_\gamma}b \\ \nonumber
K_\Lambda^2 = \frac{2 g M_1 \mu}{m_\gamma}a, 
&\hskip5mm & K_\epsilon^2 = \frac{2 g M_1 \mu}{m_\gamma}b \\ \nonumber
K_\Lambda^3 = -v_u a + \frac{v_d b}{2 v_u},
&\hskip5mm & K_\epsilon^3 = -\frac{c}{2 \mu v_u^2} 
\big(\frac{4 Det(M_H) a}{h^2 m_\gamma}-v_d v_u \mu \big)-\frac{v^2 b}{2 v_u} 
\\ \nonumber
K_\Lambda^4 = v_d a + \frac{b}{2},
&\hskip5mm & K_\epsilon^4 = \frac{h^2 \mu v_u}{4 Det(M_H)}
(4 M_1 M_2 \mu v_u - m_\gamma v_d v^2) \\ \nonumber
K_\Lambda^5 = \frac{v_R b}{2 v_u}, 
&\hskip5mm & K_\epsilon^5 = \frac{v_R c}{2 v_u} \\ \nonumber
K_\Lambda^6 = \frac{v_S b}{2 v_u}, 
&\hskip5mm & K_\epsilon^6 = \frac{c}{2 \sqrt{2} v_u v_R v_\Phi h} 
\big[ \frac{8 Det(M_H) a}{h^2 m_\gamma}+\sqrt{2} h v_\Phi v_R v_S \big] 
-\frac{2 \sqrt{2} Det(M_H) b^2}{h^3 m_\gamma v_u v_R v_\Phi} \\ \nonumber
K_\Lambda^7 = -\frac{v_\Phi b}{2 v_u},
&\hskip5mm & K_\epsilon^7 = -\frac{v_\Phi c}{2 v_u}
\end{eqnarray}
The coefficients $a$, $b$ and $c$ are defined as
\begin{eqnarray}\label{def:abc}
a=\frac{m_\gamma h^2 v_\Phi}{4 \sqrt{2} Det(M_H)} 
(-h v_R v_S+\frac{1}{2} \lambda v_\Phi^2+h_0 v_d v_u), \\ \nonumber
b=\frac{m_\gamma h^2 \mu}{4 Det(M_H)} v_u (v_u^2-v_d^2), \\ \nonumber 
c=\frac{h^2 \mu}{Det(M_H)} v_u^2 (2 M_1 M_2 \mu - m_\gamma v_d v_u).
\end{eqnarray}
$Det(M_H)$ is the determinant of the ($7,7$) matrix of the heavy 
neutral states,
\begin{equation}\label{def:detmh}
Det(M_H)=\frac{1}{16} h_0 h^2  v_\Phi^2 \big[ 4(2 M_1 M_2 \mu 
- m_\gamma v_d v_u)(-h v_R v_S+\frac{1}{2} 
\lambda v_\Phi^2+h_0 v_d v_u)-h_0 m_\gamma (v_u^2-v_d^2)^2 \big]
\end{equation}
and $v^2=v_u^2+v_d^2$. The ``photino'' mass parameter is defined 
as $m_{\gamma} = g^2M_1 +g'^2 M_2$.
Note that the $K_{\Lambda}^i$ and  $K_{\epsilon}^i$ reduce to the 
expressions of the explicit bilinear R-parity breaking model 
\cite{Hirsch:2000ef}, in the limit $M_{\Phi}\rightarrow \infty$ and 
in the limit $h,h_0\rightarrow 0$, i.e.~ $b=c=0$. 

The effective neutrino mass matrix at tree-level can then be cast into 
a very simple form 
\begin{equation}
 -(\boldsymbol{m_{\nu\nu}^{\rm eff}})_{ij} = a \Lambda_i \Lambda_j + 
     b (\epsilon_i \Lambda_j + \epsilon_j \Lambda_i) +
     c \epsilon_i \epsilon_j\,.
\label{eq:eff}
\end{equation}
Equation (\ref{eq:eff}) resembles very closely the corresponding expression 
for the explicit bilinear R-parity breaking model, once the tree-level 
and the dominant 1-loop contributions are taken into account 
\cite{Romao:1999up,Hirsch:2000ef,Diaz:2003as}. Eq. (\ref{eq:eff}) 
reduces to the tree-level expression of the explicit model 
\footnote{In the definition of the coefficient $a$ given in 
\cite{Hirsch:2004rw} there is a relative sign to the corresponding 
definition for the explicit case \cite{Hirsch:2000ef}.}
\begin{equation}
(\mathbf{m_{\nu\nu}^{\rm eff}})_{ij} = \frac{m_{\gamma}}
{4{\rm Det}\mathbf{M_{\chi^0}}} \Lambda_i \Lambda_j 
\label{eq:effexpl}
\end{equation}
in the limit $M_{\Phi}\rightarrow \infty$ and in the limit 
$h,h_0\rightarrow 0$. Different from the explicit model, however, 
the spontaneous model has in general two non-zero neutrino masses at 
tree-level. With the lightest neutrino mass zero at tree-level, the 
s-\rpv model could generate degenerate neutrinos only in regions of 
parameter space where the two tree-level neutrino masses of eq. 
(\ref{eq:eff}) are highly fine-tuned against the loop corrections. 
We will disregard this possibility in the following. 

Neutrino physics puts a number of constraints on the parameters 
$\Lambda_i$ and $\epsilon_i$. However, in the spontaneous model 
there is no a priori reason which of the terms gives the dominant 
contribution to the neutrino mass matrix, thus two possibilities 
to fit the neutrino data exist:
\begin{itemize}
\item case (c1) $\vec \Lambda$ generates the atmospheric mass scale, 
          $\vec \epsilon$ the solar mass scale
\item case (c2) $\vec \epsilon$ generates the atmospheric mass scale, 
          $\vec \Lambda$ the solar mass scale
\end{itemize}
The absolute scale of neutrino mass requires both $|\vec\Lambda|/\mu$ 
and $|\vec\epsilon|/\mu$ to be small, the exact numbers depending 
on many unknown parameters. For typical SUSY masses order 
${\cal O}(100\hskip1mm{\rm GeV})$, $|\vec\Lambda|/\mu^2 \sim 10^{-6}$--
$10^{-5}$. If some of the singlet fields are light, i.e.~have masses 
in the range of ${\cal O}(0.1-{\rm few})$ TeV, also $|\epsilon_i/\mu|$ 
can be as small as $|\vec\epsilon|/\mu \sim 10^{-6}$--$10^{-5}$. On 
the other extreme, independent of the singlet spectrum, $|\vec\epsilon|/\mu$ 
can not be larger than, say, $|\vec\epsilon|/\mu \sim 10^{-3}$, due to 
contributions from sbottom and stau loops to the neutrino mass 
matrix \cite{Romao:1999up,Hirsch:2000ef,Diaz:2003as}. 

The observed mixing angles in the neutrino sector then require certain
ratios for the parameters $\Lambda_i/\Lambda_j$ and $\epsilon_i/\epsilon_j$. 
This can be most easily understood as follows. As first observed in 
\cite{Harrison:2002er}, the so-called tri-bimaximal mixing pattern
\begin{equation}
\label{eq:UHPS}
U^{\rm HPS} =
\left(\begin{array}{cccc}
\sqrt{\frac{2}{3}} & \sqrt{\frac{1}{3}} & 0 \cr
- \frac{1}{\sqrt{6}} &  \frac{1}{\sqrt{3}} & - \frac{1}{\sqrt{2}} \cr
- \frac{1}{\sqrt{6}} &  \frac{1}{\sqrt{3}} & \frac{1}{\sqrt{2}}
\end{array}\right).
\end{equation}
is a good first-order approximation to the observed neutrino angles. In
case of hierarchical neutrinos ${\cal M}_\nu^{diag}=(0,m,M)$, where 
$m$ ($M$) stands for the solar (atmospheric) mass scale, rotating with 
$U^{\rm HPS}$ to the flavour basis leads to the following neutrino mass 
matrix
\begin{equation}
\label{eq:MHPS}
{\cal M}_\nu^{\rm HPS} =
\frac{1}{2}\left(\begin{array}{cccc}
0 &  0 &  0 \cr
0 &  M & -M \cr
0 & -M &  M
\end{array}\right) +
\frac{1}{3}\left(\begin{array}{cccc}
m &  m &  m \cr
m &  m &  m \cr
m &  m &  m
\end{array}\right).
\end{equation}
In case the coefficient $b$ in eq. (\ref{def:abc}) is exactly zero,
i.e. for $\tan\beta=1$, the model would produce a tri-bimaximal 
mixing pattern for $\Lambda_1=0,\Lambda_2=-\Lambda_3$ and 
$\epsilon_1=\epsilon_2=\epsilon_3$, in case (i). For case (ii) the 
conditions on $\Lambda_i$ should be exchanged with the conditions for 
the $\epsilon_i$ and vice versa. 

In reality, since $\tan\beta\ne 1$ in general, neither is $b$ exactly 
zero, nor need the neutrino mixing angles be exactly those of eq. 
(\ref{eq:UHPS}). One then finds certain allowed ranges for ratios of 
the $\Lambda_i$ and $\epsilon_i$. In case (i) one gets approximately 
\begin{eqnarray}\label{fitnucase1}
\Big(\frac{\Lambda_1}{\sqrt{\Lambda_2^2+\Lambda_3^2}}\Big)^2 
\simeq \tan^2\theta_{\rm R},
\\ \nonumber
\Big(\frac{\Lambda_2}{\Lambda_3}\Big)^2 \simeq \tan^2\theta_{\rm Atm},
\\ \nonumber
\Big(\frac{{\tilde\epsilon_1}}{{\tilde\epsilon_2}}\Big)^2 
\simeq \tan^2\theta_{\odot}.
\end{eqnarray}
Here, ${\tilde\epsilon} = U_{\nu}^T\cdot{\vec\epsilon}$ with 
$(U_{\nu})^T$ being the matrix which diagonalizes the ($3,3$) 
effective neutrino mass matrix. In case (i) $(U_{\nu})^T$ is 
(very) approximately given by
\begin{equation}\label{def:tildeeps}
{\tilde\epsilon} = 
\left(\begin{array}{cccc}
\frac{\sqrt{\Lambda_2^2 + \Lambda_3^2}}{|\vec\Lambda|} &  
-\frac{\Lambda_1\Lambda_2}{\sqrt{\Lambda_2^2 + \Lambda_3^2}|\vec\Lambda|} &  
-\frac{\Lambda_1\Lambda_3}{\sqrt{\Lambda_2^2 + \Lambda_3^2}|\vec\Lambda|} \cr 
 0 &  
\frac{\Lambda_3}{\sqrt{\Lambda_2^2 + \Lambda_3^2}} 
& -\frac{\Lambda_2}{\sqrt{\Lambda_2^2 + \Lambda_3^2}}  \cr 
 \frac{\Lambda_1}{|\vec\Lambda|} & 
 \frac{\Lambda_2}{|\vec\Lambda|} & 
 \frac{\Lambda_3}{|\vec\Lambda|} \cr
\end{array}\right)\cdot
{\vec \epsilon}
\end{equation}
Note that $ U_{\Lambda}^T$ is the matrix which diagonalizes only 
the part of the effective neutrino mass matrix proportional to 
$\Lambda_i\Lambda_j$. Again, for the case (ii) replace 
$\Lambda_i\leftrightarrow \epsilon_i$ in all expressions.

\subsection{Approximated couplings}
\label{sec:cpl}

With R-parity violated the lightest supersymmetric particle decays. 
Here we list the most important couplings of the lightest neutralino 
in the seesaw approximation. In the numerical calculation discussed 
in the next sections, we always diagonalize all mass matrices exactly 
and obtain the exact couplings. For the understanding of the main 
qualitative features of the LSP decays, however, the approximated 
couplings listed below will be very helpful. 

We define the ``rotated'' quantities:
\begin{eqnarray}\label{def:rot}
\tilde{x}_i \equiv \big( U_\nu \big)_{ik}^T x_k, & \hskip5mm & 
\tilde{y}_{ij} \equiv \big( U_\nu \big)_{ik}^T y_{kj}.
\end{eqnarray}

$\tilde{\chi}_1^0-W^{\pm}-l^{\mp}_i$ couplings are found from the 
general expressions for the $\tilde{\chi}^0-W^{\pm}-{\tilde \chi}^{\mp}$ 
vertices 
\begin{equation}\label{eq:cnw}
\mathcal{L}=\bar{\chi}_i^- \gamma^{\mu} 
\big( O_{Lij}^{cnw} P_L + O_{Rij}^{cnw} P_R \big) \chi_j^0 W_\mu^- 
+ \bar{\chi}_i^0 \gamma^{\mu} \big( O_{Lij}^{ncw} P_L + O_{Rij}^{ncw} 
P_R \big) \chi_j^- W_\mu^+
\end{equation}
as 
\begin{eqnarray}\label{eq:cnwdef}
O_{Li1}^{cnw}& =& \frac{g}{\sqrt{2}} \big[ 
\frac{g N_{12} \Lambda_i}{Det_+}-\big( \frac{\epsilon_i}{\mu}
+\frac{g^2 v_u \Lambda_i}{2 \mu Det_+} \big)N_{13}
-\displaystyle\sum_{k=1}^{7} N_{1k} \xi_{ik} \big], \\ \nonumber
O_{Ri1}^{cnw} & = & \frac{1}{2} g (h_E)_{ii} \frac{v_d}{Det_+} 
\big[ \frac{g v_d N_{12} + M_2 N_{14}}{\mu} \epsilon_i+
\frac{g(2 \mu^2 +g^2 v_u v_d)N_{12}+g^2 v_u(\mu+M_2)N_{14}}{2 \mu Det_+} 
\Lambda_i \big].
\end{eqnarray}
$Det_+$ is the determinant of the MSSM chargino mass matrix. Here, 
\begin{eqnarray}\label{eq:cnwLR}
O_{Li1}^{ncw}& =& \big(O_{Li1}^{cnw}\big)^*, \\ \nonumber
O_{Ri1}^{ncw}& =& \big(O_{Ri1}^{cnw}\big)^*.
\end{eqnarray}
The Lagrangian for $\tilde{\chi}_i^0-\tilde{\chi}_j^0-Z$
\begin{equation}
\mathcal{L}=\frac{1}{2} \bar{\chi}_i^0 \gamma^{\mu} \big( O_{Lij}^{nnz} 
P_L + O_{Rij}^{nnz} P_R \big) \chi_j^0 Z_\mu
\end{equation}
gives for $\tilde{\chi}_1^0-\nu_i-Z$
\begin{eqnarray}\label{eq:cnndef}
O_{Li1}^{nnz}& =& -\frac{g}{2 \cos \theta_W} 
\big[\tilde{\xi}_{i1} N_{11}+\tilde{\xi}_{i2} N_{12} 
+ 2 \tilde{\xi}_{i4} N_{14} +\tilde{\xi}_{i5} N_{15}
+\tilde{\xi}_{i6} N_{16}+\tilde{\xi}_{i7} N_{17}\big], \\ \nonumber
O_{Ri1}^{nnz}& =& -\big( O_{Li1}^{nnz} \big)^*.
\end{eqnarray}
The most important difference to the explicit R-parity violating 
models comes from the coupling ${\chi}_i^0-{\chi}_j^0-P^0_k$
\begin{equation}\label{defnnp}
\mathcal{L}=\frac{1}{2} \bar{\chi}_i^0 
\big( O_{Lijk}^{nnp} P_L + O_{Rijk}^{nnp} P_R \big) \chi_j^0 P_k^0,
\end{equation}
with
\begin{eqnarray}
O_{Li1J}^{nnp}& =& R_{Jm}^p O_{Li1m}^{' nnp}, \\ \nonumber
O_{Ri1J}^{nnp}& = &\big(O_{Li1J}^{nnp}\big)^*.
\end{eqnarray}
Because the spontaneous breaking of lepton number produces 
a massless pseudo-scalar, eq.(\ref{defnnp}) leads to a 
coupling $\tilde{\chi}_1^0-J-\nu_i$, i.e~ a new invisible 
decay channel for the lightest neutralino. For $v_L \ll v_R,v_S$ 
one can find an approximation to the Majoron which, in leading 
order, is given by
\begin{equation}\label{smplstmaj}
R_{Jm}^p \simeq 
\big(0,0,\frac{v_{L k}}{V},0,\frac{v_S}{V},-\frac{v_R}{V}\big). 
\end{equation}
Here, $V=\sqrt{v_R^2+v_S^2}$ and terms of order $\frac{v_L^2}{V v}$ 
have been neglected. The ``unrotated'' couplings $O_{Li1m}^{' nnp}$ 
are
\begin{eqnarray}\label{defOnnpur}
O_{Li1\tilde{L}_k^0}^{' nnp} & =&-\frac{i}{2} \big(U_\nu\big)_{ki}
(g' N_{11}-g N_{12}), \\ \nonumber
O_{Li1\tilde{S}}^{' nnp}& =& \frac{i}{\sqrt{2}} h 
\big( \tilde{\xi}_{i5} N_{17}+\tilde{\xi}_{i7} N_{15} \big), \\ \nonumber
O_{Li1\tilde{\nu}^c}^{' nnp} &= & -i \frac{\tilde{\epsilon}_i}{v_R} N_{14}
+\frac{i}{\sqrt{2}} h \big(\tilde{\xi}_{i6} N_{17}+\tilde{\xi}_{i7} N_{16} 
\big).
\end{eqnarray}
In the limit $v_R, v_S \rightarrow \infty$ one can derive a very simple 
approximation formula for $O_{\tilde\chi^0_1\nu_kJ}$. It s given by
\footnote{We correct here a misprint in \cite{Hirsch:2006di}.}
\begin{equation}
\label{eq:majcl}
|O_{\tilde\chi^0_1\nu_kJ}| \simeq  - \frac{{\tilde \epsilon}_k}{V}N_{14} +
\frac{{\tilde v}_{L_k}}{2 V}(g' N_{11} - g N_{12})
+ h.O.
\end{equation}
Eq. (\ref{eq:majcl}) serves to show that for constant ${\tilde \epsilon}$ 
and ${\tilde v}_{L}$, $O_{\tilde\chi^0_1\nu_kJ} \rightarrow 0$ as $v_R$ 
goes to infinity. This is as expected, since for $v_R\rightarrow\infty$ 
the spontaneous model approaches the explicit bilinear model. Note, that 
only the presence of the field $\widehat \nu^c$ is essential for the 
coupling Eq.~(\ref{eq:majcl}). If $\widehat S$ is absent, replace 
$V \rightarrow v_R$. 

In addition to the Majoron in considerable parts of the parameter 
space one also finds a rather light singlet scalar, called the 
``scalar partner'' of the Majoron in \cite{Hirsch:2005wd}, $S_J$. 
From the Lagrangian
\begin{equation}
\mathcal{L}=\frac{1}{2} \bar{\chi}_i^0 \big( O_{Lijk}^{nns} P_L 
+ O_{Rijk}^{nns} P_R \big) \chi_j^0 S_k^0,
\end{equation}
one finds the coupling $\tilde{\chi}_1^0-S_J-\nu_i$ as
\begin{eqnarray}
O_{Li1S_J}^{nns}& =& R_{S_Jk}^s O_{Li1k}^{' nns}, \\ \nonumber
O_{Ri1S_J}^{nns}& =& \big(O_{Li1S_J}^{nns}\big)^*.
\end{eqnarray}
Different from the Majoron, however, there is no simple analytical 
approximation for $R_{S_J}$. We write symbolically
\begin{equation}\label{defsj}
R_{S_Jk}^s = \big(R_{S_J H_d},R_{S_J H_u},R_{S_J \tilde{L}_k^0},
R_{S_J \Phi},R_{S_J \tilde{S}},R_{S_J \tilde{\nu}^c} \big) ,
\end{equation}
and define unrotated couplings by
\begin{eqnarray}\label{defonnsur}
O_{Li1H_d}^{' nns}& =& \frac{1}{2} \big[ 
(g \tilde{\xi}_{i2}-g' \tilde{\xi}_{i1})N_{13} 
+(g N_{12}-g' N_{11}) \tilde{\xi}_{i3} 
-\sqrt{2} h_0 (\tilde{\xi}_{i7} N_{14}
+\tilde{\xi}_{i4} N_{17})\big], \\ \nonumber 
O_{Li1H_u}^{' nns}& =& \frac{1}{2} \big[ (g' \tilde{\xi}_{i1}
-g \tilde{\xi}_{i2})N_{14} +(g' N_{11}-g N_{12}) \tilde{\xi}_{i4} 
-\sqrt{2} h_0 (\tilde{\xi}_{i7} N_{13}+\tilde{\xi}_{i3} N_{17})
-\sqrt{2} \frac{\tilde{\epsilon}_i}{v_R} N_{15}\big], \\ \nonumber
O_{Li1\tilde{L}_k^0}^{' nns} &=& \frac{1}{2} \big(U_\nu\big)_{ki}
(g' N_{11}-g N_{12}), \\ \nonumber 
O_{Li1\Phi}^{' nns} &= & \frac{1}{\sqrt{2}} \big[-h_0 (\tilde{\xi}_{i3} 
N_{14}+\tilde{\xi}_{i4} N_{13})+h(\tilde{\xi}_{i6} N_{15}
+\tilde{\xi}_{i5} N_{16})+\lambda \tilde{\xi}_{i7} N_{17}\big], \\ \nonumber 
O_{Li1\tilde{S}}^{' nns} &= & \frac{1}{\sqrt{2}} h 
\big( \tilde{\xi}_{i5} N_{17}+\tilde{\xi}_{i7} N_{15} \big), \\ \nonumber 
O_{Li1\tilde{\nu}^c}^{' nns} &= & -\frac{\tilde{\epsilon}_i}{v_R} N_{14}
+\frac{1}{\sqrt{2}} h \big(\tilde{\xi}_{i6} N_{17}
+\tilde{\xi}_{i7} N_{16} \big).
\end{eqnarray}
As eqs (\ref{defonnsur}) shows, $\tilde{\chi}_1^0\to S_J + \nu_i$ has a 
partial decay width similar in size to the decay $\tilde{\chi}_1^0\to J +
\nu_i$, as soon as kinematically allowed. Since, on the other hand, $S_J$ 
decays practically always with a branching ratio close to 100 \% 
into two Majorons, $\tilde{\chi}_1^0\to S_J + \nu_i$ gives in general a 
sizeable contribution to the invisible width of the neutralino. 

Finally, we give also the coupling $\tilde{\chi}_i^0-J-\tilde{\chi}_j^0$, 
for the case of two heavy neutralinos. Here, 
\begin{equation}
O_{LijJ}^{nnp}=-\frac{i}{\sqrt{2}} \frac{h}{V} \big[ 
v_S (N_{j7} N_{i5}+N_{i7} N_{j5})-v_R (N_{j7} N_{i6}+N_{i7} N_{j6}) \big].
\end{equation}

\section{LSP production and decays}
\label{sec:pheno}

In this section we discuss the phenomenology of a neutralino LSP in 
s-\rpv at future colliders. We do not attempt to do an exhaustive 
study of the (quite large) parameter space of the model. Instead 
we will focus on the most important qualitative differences between 
s-\rpv, the previously studied case of explicit bilinear \rpv 
\cite{Mukhopadhyaya:1998xj,Choi:1999tq,Porod:2000hv,deCampos:2007bn}
 and the MSSM. All numerical 
results shown below have been obtained using the program package 
SPheno \cite{Porod:2003um}, extended to include the new singlet 
superfields $\widehat\nu^c$, $\widehat S$ and $\widehat\Phi$. 

Unless mentioned otherwise, we have always chosen the \rpv parameters in 
such a way that solar and atmospheric neutrino data \cite{Maltoni:2004ei} 
are fitted in the correct way. The numerical procedure to fit neutrino 
masses is the following. Compared to the MSSM we have a number of 
new parameters. For the superpotential of eq. (\ref{eq:Wsuppot}) these 
are $h_0$, $h$ and $\lambda$, as well as the neutrino Yukawas $h_{\nu}^i$. 
In addition, there are in principle also the soft SUSY breaking terms, 
which generate non-zero vevs, $v_R$, $v_S$, $v_{\Phi}$ and $v_{L_i}$ 
for ${\tilde\nu}^c$, ${\tilde S}$, $\Phi$ and ${\tilde\nu}_i$, respectively. 
We trade the unknown soft parameters for the vevs. For any random 
choice of MSSM parameters, we can reproduce the ``correct'' MSSM value 
of $\mu$ for a random value of $v_{\Phi}$, by appropriate choice of 
$h_0$. For any random set of $h$, $\lambda$, $v_S$ and $v_R$, we can 
then calculate those values of $h_{\nu}^i$ and $v_{L_i}$, using 
eq. (\ref{eq:eff}), such that the corresponding $\epsilon_i$ and 
$\Lambda_i$ give correct neutrino masses and mixing angles. There 
are two options, how neutrino data can be fitted, i.e. the cases (c1) 
and (c2), defined in section \ref{sec:neutrino-masses}. We discuss 
the differences between these two possibilities below. 

In the following we will study only two 'limiting' cases, which we 
consider to be the simplest possibilities to realize within the parameter 
space of the model: (a) a bino-like LSP and (b) a singlino LSP. We note, 
however, that theoretically also other possibilities exist at least in 
some limited parts of parameter space. For example, one could also have 
that the phino, ${\tilde\Phi}$, is the lightest $R_p$ odd state. However, 
with the superpotential of eq. (\ref{eq:Wsuppot}), for any given value 
of $\mu$, $v_{\Phi}$ has a minimum value. Since the product 
$\lambda v_{\Phi}$ also determines approximately the phino mass, a very 
light phino requires a certain hierarchy $\lambda \ll h_0,h$, which might 
be considered to be a rather special case. Also in mSugra in the region 
where $m_0$ is large one can find points in which $\mu \sim M_1$ and the 
lightest (MSSM) neutralino has a significant higgsino component. Since 
both, a higgsino as well as a phino LSP show some differences in 
phenomenology compared to the bino and singlino LSPs discussed here, 
we plan to study higgsino and phino LSPs in a future publication.

\subsection{Production}

Since neutrino physics requires that the R-parity violating parameters
are small, supersymmetric production cross sections are very similar
to the corresponding MSSM values, see for example
\cite{AguilarSaavedra:2005pw} and references therein. Over most of the
MSSM parameter space one expects that mainly gluinos and squarks are
directly produced at the LHC and that the lightest neutralinos appear as
the ``final'' decay products at the end of possibly long decay chains
of sparticles. In addition charginos, neutralinos and sleptons
can be produced directly via Drell-Yan processes provided that they are
relatively light.

Cross sections for direct production of singlinos are 
always negligible. There are essentially two possibilities how singlinos
can be produced in cascade decays.
Firstly, a somewhat exotic chance to produce singlinos occurs if 
at least one of the MSSM Higgsinos is heavier than $\tilde\Phi$ {\em and 
both} $h_0$ and $h$ are large. In this case ${\cal S}_i$ appear in  
decay chains such as $\tilde H_{u,d}\to \tilde\Phi + \,X_1 \to {\cal S}+\,X_2$,
where $X_i$ denotes the additionally produced particles.
Secondly, there is the possiblity that singlinos are the LSPs. 
Squarks and gluinos will then decay fast to the NLSP, which then decays 
to ${\cal S}$. A typical decay chain might be ${\tilde q} \to q 
+ {\tilde B} \to q + {\cal S}_{1,2} + J$. Other NLSPs such as, 
${\tilde\tau}_1$ will decay mainly via ${\tilde\tau}_1 \to {\cal S}_{1,2} 
+ \tau$, i.e. again ending up in singlinos. The total number of singlino 
events therefore will be simply approximately equal to the number of 
SUSY events for singlino LSPs.

\subsection{Decays}

\begin{figure}[htbp]
\begin{center}
\vspace{5mm}
\includegraphics[width=80mm,height=60mm]{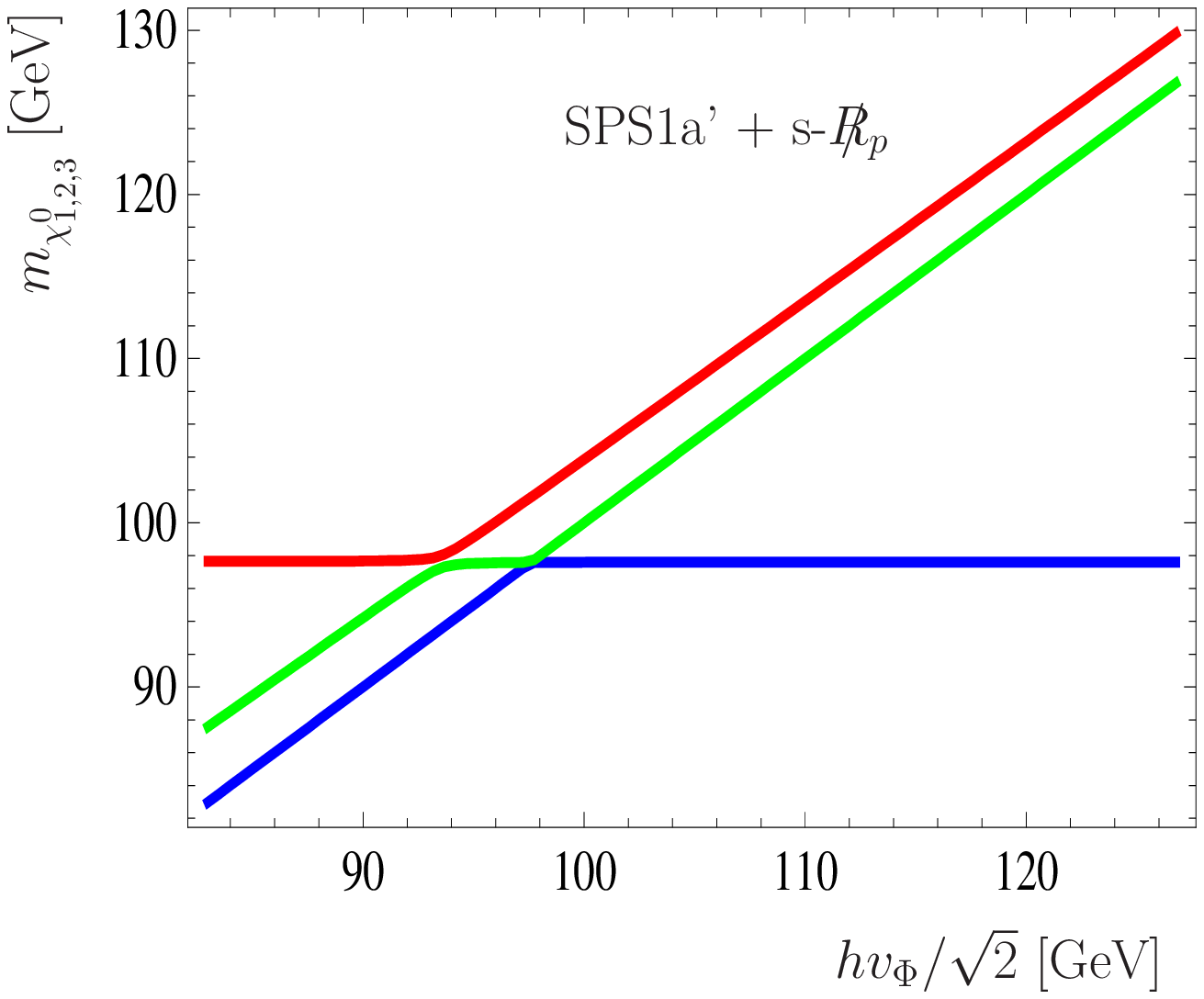}
\includegraphics[width=80mm,height=60mm]{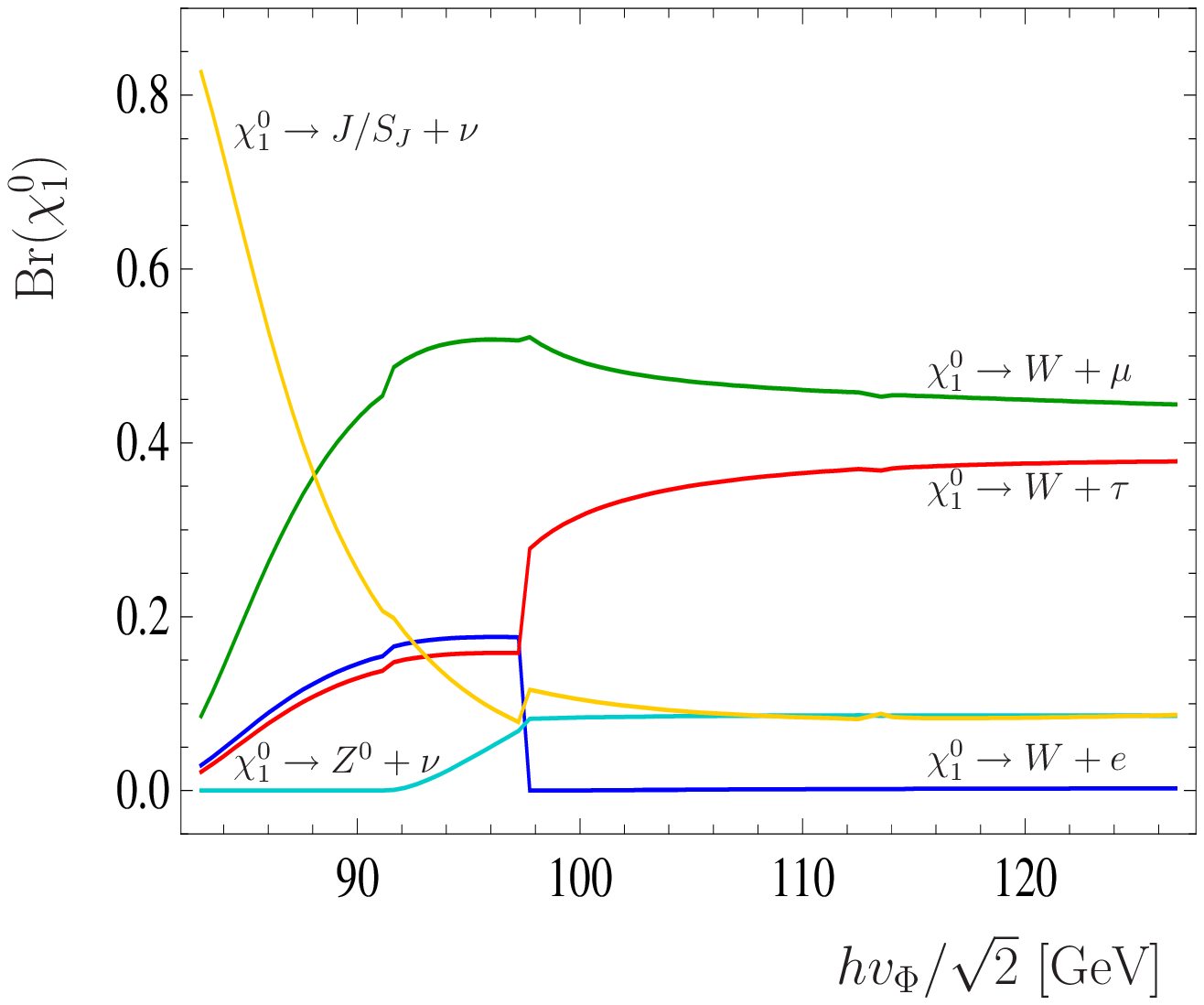}
\end{center}

\caption{Masses of the three lightest neutralinos (left) and branching 
ratios for the most important decay modes of the lightest state (right) 
versus $\frac{1}{\sqrt{2}}hv_{\Phi}$ for a specific, but typical example 
point. The MSSM parameters have been adjusted such that the sparticle 
spectrum of the standard point SPS1a' is approximately reproduced. The 
singlet parameters have been chosen randomly, $v_R=v_S=1$ TeV, 
${\vec\epsilon}$ and $\vec\Lambda$ have been fitted to neutrino data, 
such that $\vec\Lambda$ generates the atmospheric scale and ${\vec\epsilon}$ 
the solar scale. For a detailed discussion see text.}
\label{fig:Exa_mass_br}
\end{figure}

Here we will discuss the main decay modes of bino and singlino LSPs. 
We will first discuss the parameter range, where $m_{\chi^0_1} 
\ge m_{W^{\pm}}$, such that two-body decays of $\chi^0_1$ to gauge bosons 
are kinematically allowed. 
Fig. (\ref{fig:Exa_mass_br}) shows an example of the three lightest 
neutralino mass eigenvalues (left) and the main decay modes of $\chi^0_1$ 
(right) as a function $\frac{hv_{\Phi}}{\sqrt{2}}$ for fixed values of all  
other parameters. This point has been constructed in such a way, that 
the MSSM part of the spectrum, all production cross sections and all 
decay branching ratios, apart from the lightest neutralino decays, match 
very closely the mSugra standard point SPS1a' \cite{AguilarSaavedra:2005pw}. 
Here, $v_R=v_S=1$ TeV has been chosen as an arbitrary, but typical example. 

\begin{figure}[htbp]
\begin{center}
\vspace{5mm}
\includegraphics[width=80mm,height=60mm]{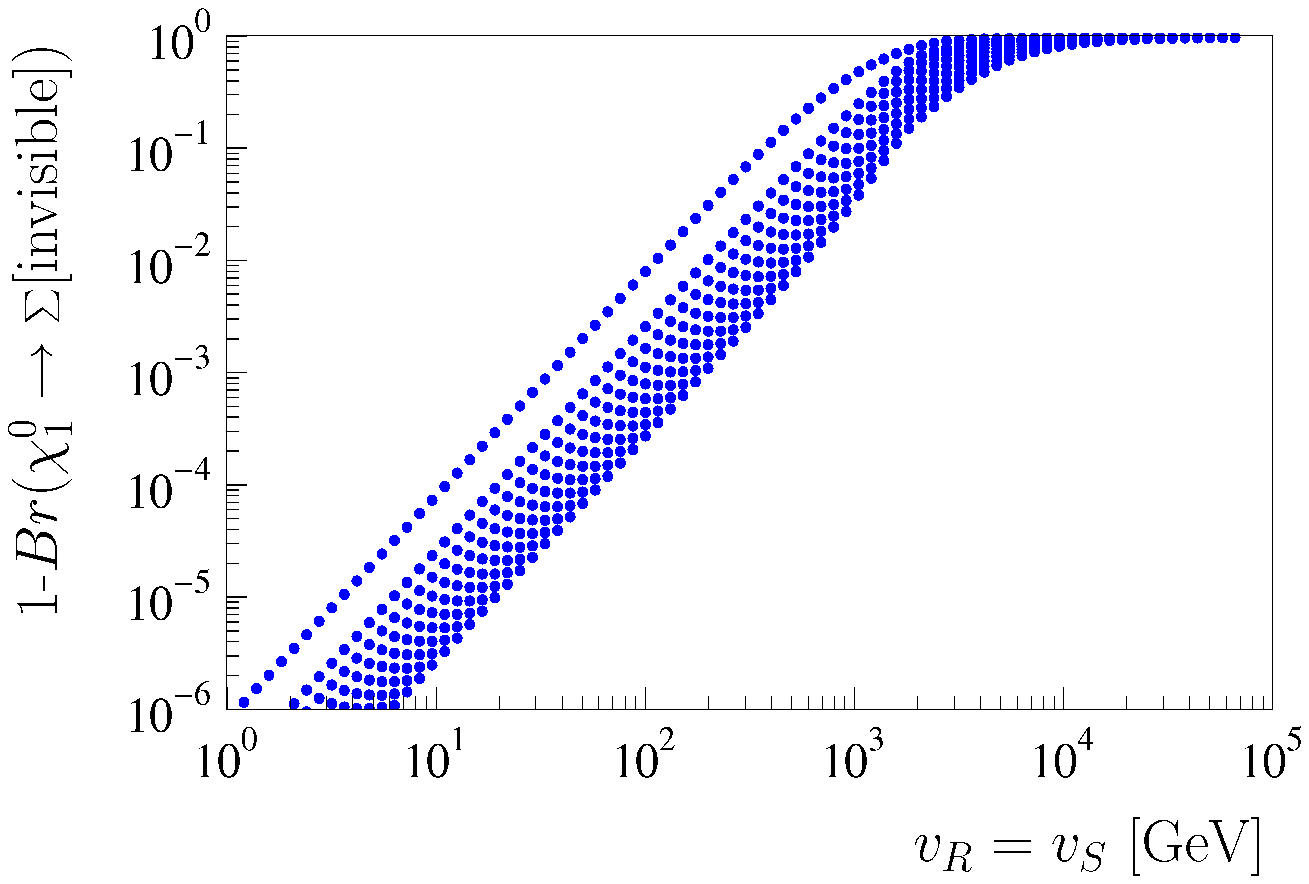}
\includegraphics[width=80mm,height=60mm]{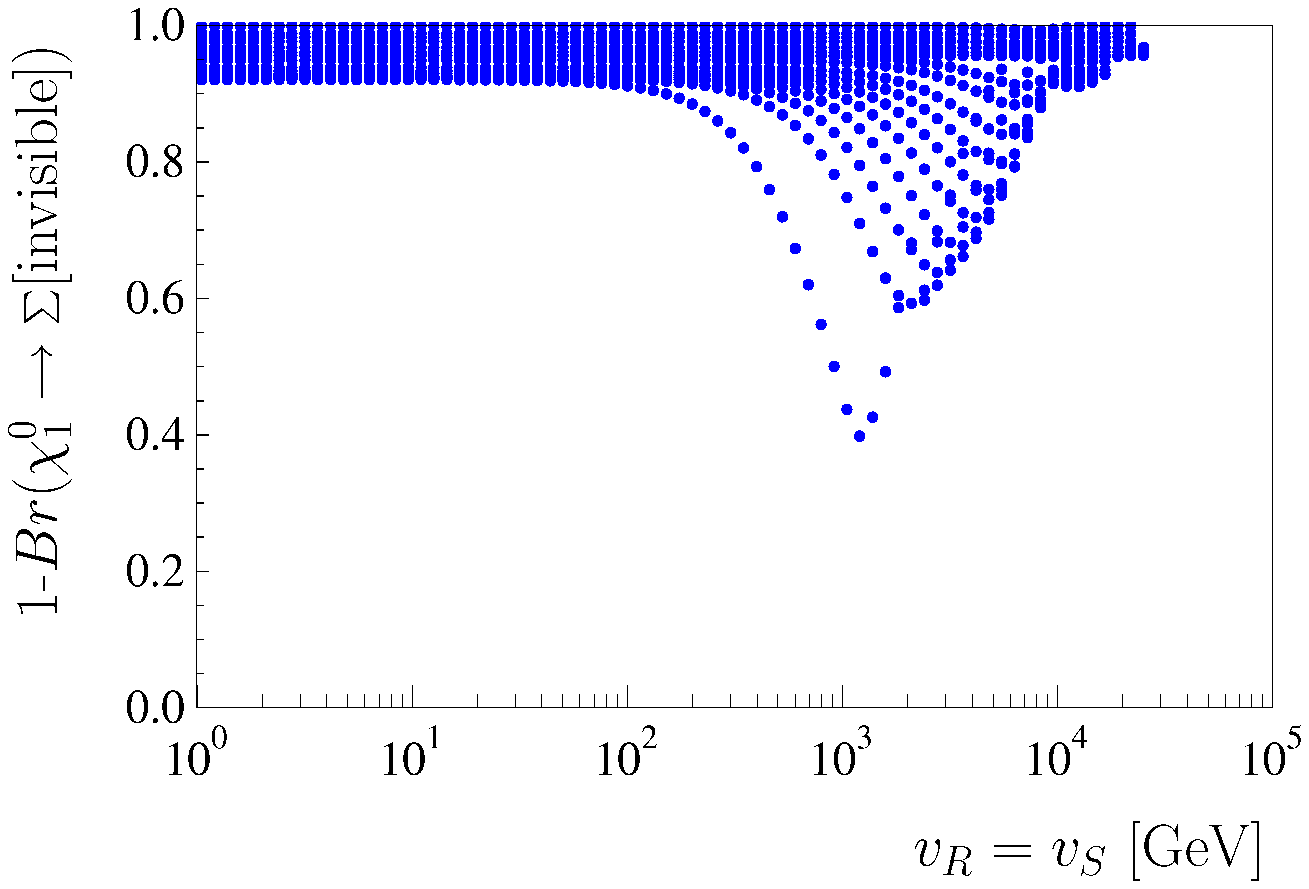}
\end{center}

\caption{Sum over all at least partially visible decay modes 
of the lightest neutralino versus $v_R$ in GeV, for a set of 
$v_{\Phi}$ values $v_{\Phi} = 10$--40 TeV for the mSUGRA parameter 
point $m_0=280$ GeV, $m_{1/2}=250$ GeV, $\tan\beta=10$, $A_0=-500$ GeV 
and $sgn(\mu)=+$. To the left $\chi_1^0 \simeq {\tilde B}$; to the right 
$\chi_1^0 \simeq {\cal S}$. The plot demonstrates that the branching 
ratio into ${\tilde B} \to J + \nu$ does depend strongly on the value 
of $v_R$ and to a minor extend on $v_{\Phi}$. Lowering $v_R$ one can 
get branching ratios for the invisible decay of the ${\tilde B}$ very close 
to 100 \%, thus a very MSSM-like phenomenology. The right plot demonstrates 
that such a possibility does not exist in the case of an ${\cal S}$ LSP.}
\label{fig:Inv_vr}
\end{figure}

The left part of fig. (\ref{fig:Exa_mass_br}) shows how the quasi-Dirac 
pair ${\cal S}_{1,2}$ evolves as a function of $\frac{hv_{\Phi}}{\sqrt{2}}$. 
For low values (i.e. $\lsim M_1$)of this parameter combination ${\cal S}_{1}$ 
is the LSP, for large 
values a ${\tilde B}$ is the LSP. The right side of the figure shows 
the final states with the largest branching ratios. For low values of 
the LSP mass, $J/S_J+\nu$ is usually the most important, i.e. there is 
a sizeable decay to invisible final states, even for a relatively high 
$v_R$, see also the discussion for fig. (\ref{fig:Inv_vr}). Next in 
importance are the final states involving $W^{\pm}$ and charged leptons. 
Note, that the model predicts 
\begin{equation}\label{ZtoW}
\frac{\sum_i Br(\chi_1^0\to Z^0 + \nu_i)}
     {2 \sum_i Br(\chi_1^0\to W^{+} + l_i^{-})} 
     \simeq \frac{g}{4 \cos^2\theta_W}
\end{equation}
with $g$ being a phase space correction factor, with $g \to 1$ 
in the limit $m_{\chi^0_1}\to\infty$ \cite{Hirsch:2005ag}. Equation 
(\ref{ZtoW}) can be understood with the help of the approximative couplings 
eq. (\ref{eq:cnwdef}) and eq. (\ref{eq:cnndef}). The relative size 
of the branching ratios for the final states $W+e$, $W+\mu$ and $W+\tau$ 
depends on both, (a) the nature of the LSP and (b) the fit to the 
neutrino data. We will discuss this important feature in more detail 
in section \ref{sec:corr}.

Generally, for $m_{\chi^0_1}\ge m_{W^\pm}$ three-body final states of 
the neutralino decay are less important than the two-body decays shown 
in fig. (\ref{fig:Exa_mass_br}). Especially one expects that the final 
state $\nu b{\bar b}$ has a smaller branching than in the case of 
explicit \rpv \cite{Porod:2000hv}. This is essentially due to the fact, 
that $|\vec\epsilon|/\mu$ is smaller in s-\rpv with a ``light'' singlet 
spectrum than in a model with explicit bilinear \rpv, see the discussion 
in section \ref{sec:neutrino-masses}. A smaller $|\vec\epsilon|/\mu$ 
leads to smaller couplings between $\chi^0_1-l-{\tilde l}$, 
$\chi^0_1-q-{\tilde q}$ and especially $\chi^0_1-\nu-h^0$, see also 
couplings in \cite{Porod:2000hv}. We have checked numerically, that 
Br($\chi^0_1\to\nu + h^0$), if kinematically open, is typically below 
$1 \%$ for singlets in the ${\cal O}({\rm TeV})$ range.

\begin{figure}[htbp]
\begin{center}
\vspace{5mm}
\includegraphics[width=80mm,height=60mm]{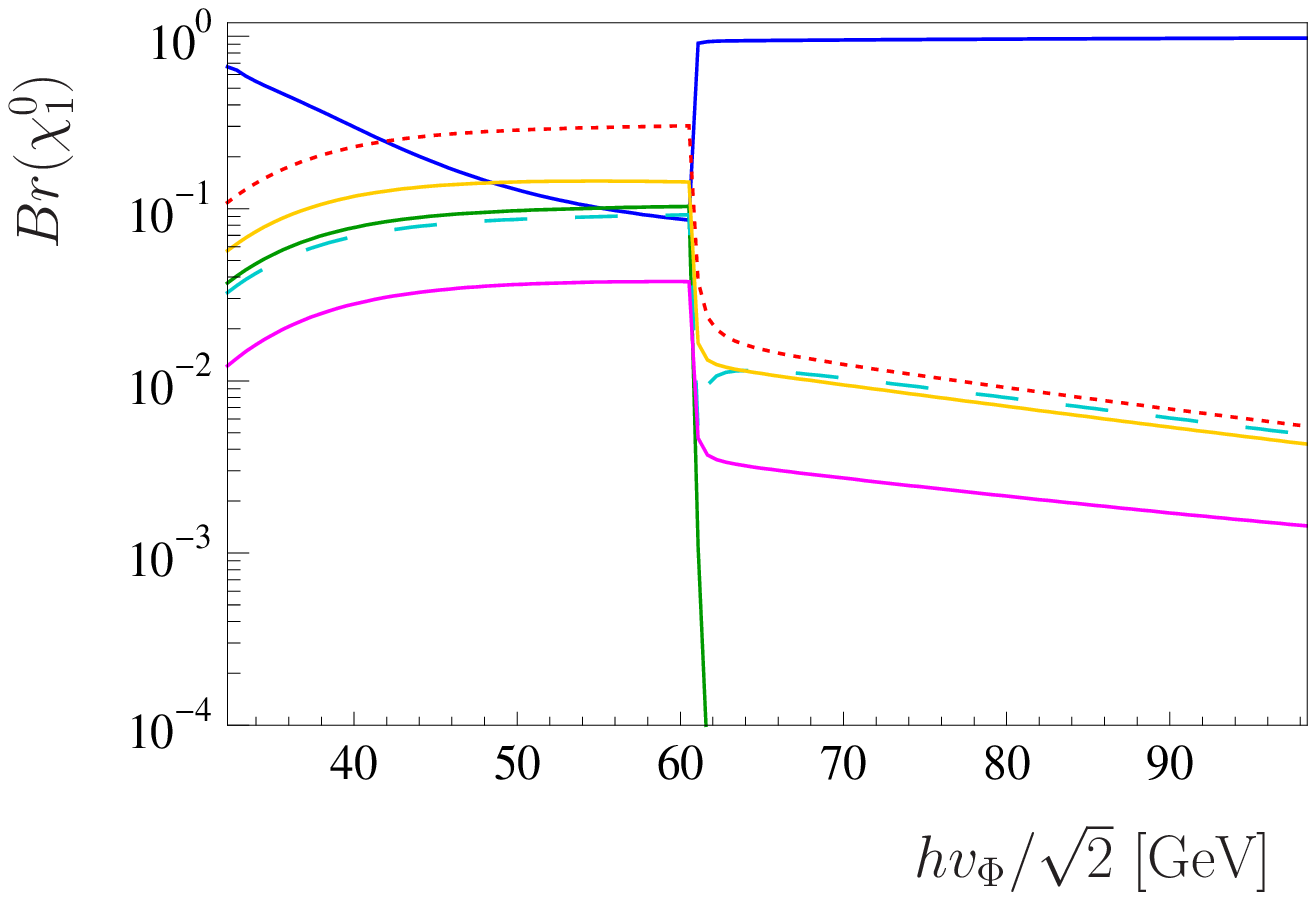}
\includegraphics[width=80mm,height=60mm]{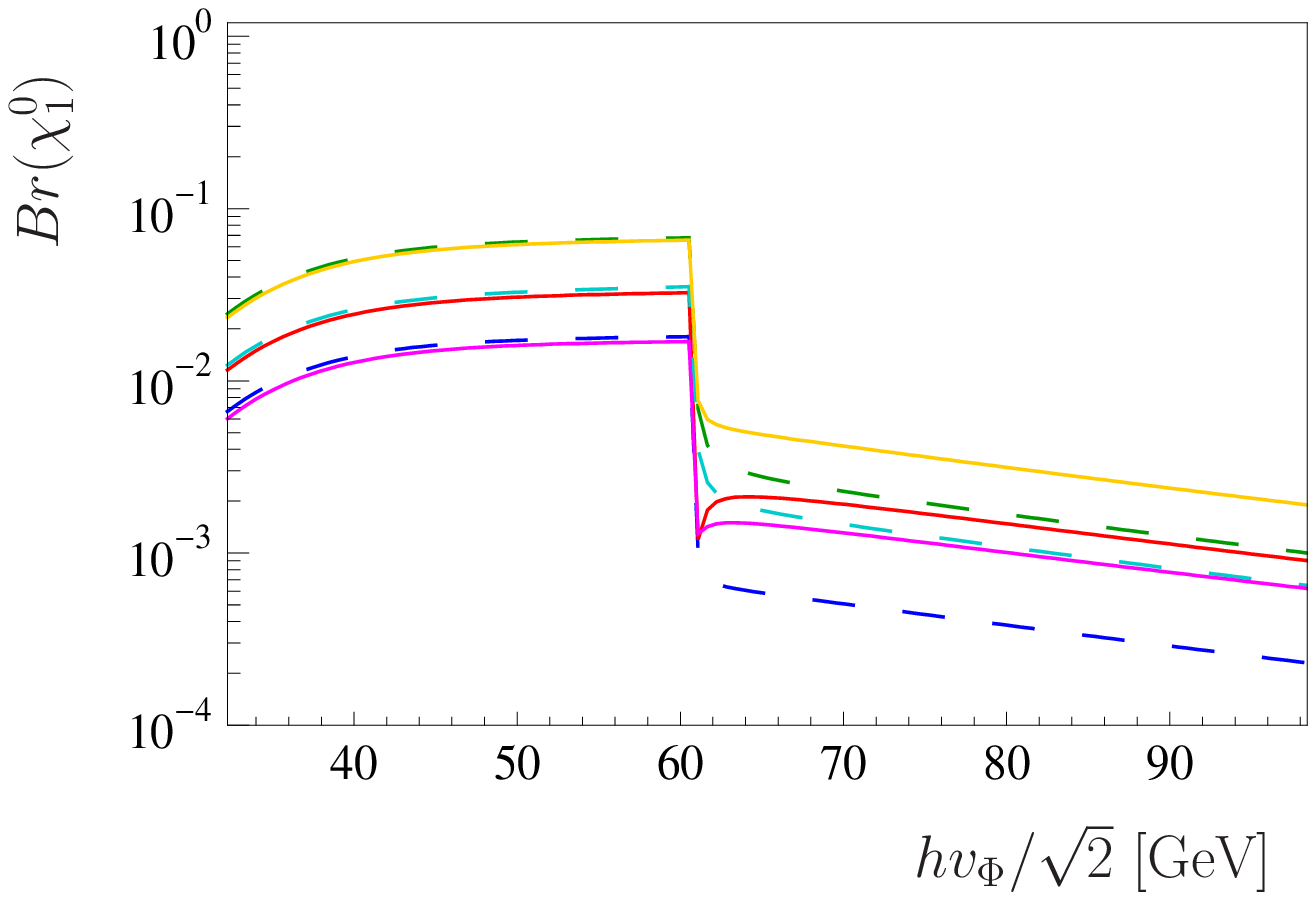}
\end{center}

\caption{Branching ratios for the most important decay modes of the 
lightest neutralino state versus $\frac{1}{\sqrt{2}}hv_{\Phi}$ for a 
specific, but typical example point. The MSSM parameters have been 
adjusted such that the sparticle spectrum of the standard point SU4 
is approximately reproduced. The singlet parameters have been chosen 
randomly, $v_R=v_S=1$ TeV, ${\vec\epsilon}$ and $\vec\Lambda$ have 
been fitted to neutrino data, such that $\vec\Lambda$ generates the 
atmospheric scale and ${\vec\epsilon}$ the solar scale. The different 
final states are as follows. In the left figure, as ordered on the right 
side, from top to bottom the lines are 
Br($\chi^0_1\to [{\rm invisible}]$) (full line, blue), 
Br($\chi^0_1\to \mu qq'$) (short-dashed, red), 
Br($\chi^0_1\to \tau qq'$) (large-dashed, light blue), 
Br($\chi^0_1\to \nu q{\bar q}$) (full, yellow),  
Br($\chi^0_1\to \nu b{\bar b}$) (full, pink) and 
Br($\chi^0_1\to e qq'$) (full, green). In the right 
figure, purely leptonic modes, from top to bottom (on the right 
side): 
Br($\chi^0_1\to \nu \mu\tau$) (full, yellow), 
Br($\chi^0_1\to \nu e \mu$) (dashed, green), 
Br($\chi^0_1\to \nu e \tau$) (full, red), 
Br($\chi^0_1\to \nu \mu \mu$) (dashed, light blue),
Br($\chi^0_1\to \nu \tau \tau$) (full, pink) and
Br($\chi^0_1\to \nu e e$) (dashed, darker blue). 
For a detailed discussion see text.}
\label{fig:Exa_br_low}
\end{figure}

For the case of $m_{\chi^0_1}\le m_{W^\pm}$ fig. (\ref{fig:Exa_br_low}) 
shows an example for the most important final states of the lightest 
neutralino decay as a function of $\frac{hv_{\Phi}}{\sqrt{2}}$. As in 
the fig. (\ref{fig:Exa_mass_br}) to the left the lightest neutralino 
is a singlino, to the right of the ``transition'' region the lightest 
neutralino is a bino. Note that the point SU4 \cite{atlas:su4} 
produces a bino mass of approximately $m_{\tilde B} \simeq 60$ GeV, 
thus the only two body decay modes which are kinematically allowed 
are $J+\nu$ and - very often, but not always - $S_J+\nu$. One observes 
that these invisible decay modes have typically a larger branching 
ratio than in the case $m_{\chi^0_1}\ge m_{W^\pm}$ shown in fig. 
(\ref{fig:Exa_mass_br}). This fact is essentially due to the propagator 
and phase space suppression factors for three body decays. For a bino 
LSP the invisible decay has the largest branching fraction. Semileptonic 
modes are next important with typically $l_i qq'$ being larger than 
$\nu q{\bar q}$. It is interesting to note, that in the purely leptonic 
decays, lepton flavour violating final states such as $\mu\tau$ have 
branching ratios typically as large or larger than the corresponding 
charged lepton flavour diagonal decays ($\mu\mu$ and $\tau\tau$). These 
large flavour off-diagonal decays can be traced to the fact that 
neutrino physics requires two large mixing angles. The branching ratios 
shown in fig. (\ref{fig:Exa_br_low}) should be understood only as 
representative examples - not as firm predictions. Especially for the 
case of a bino LSP, the partial width to the final state 
$J+\nu$, i.e. invisible final state, can vary by several orders of 
magnitude, see the discussion below. The predictions for relative 
ratios of the different (partially or completely) visible final states 
is much tighter fixed, because these final states correlate with neutrino 
physics, as we discuss in section \ref{sec:corr}.

\begin{figure}[htbp]
\begin{center}
\vspace{5mm}
\includegraphics[width=80mm,height=60mm]{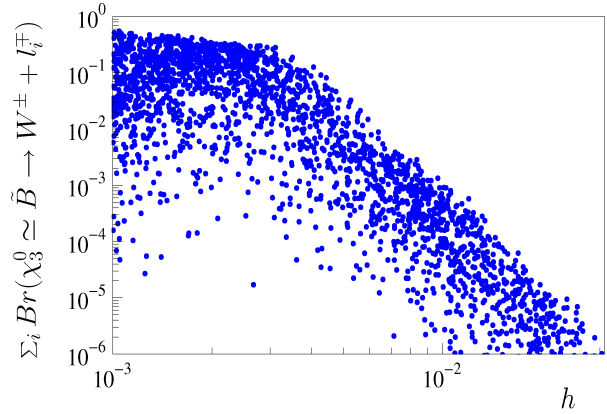}
\end{center}

\caption{Sum over 
$\sum_i Br(\chi^0_3 \simeq {\tilde B} \rightarrow  W^{\pm} + l_i^{\mp})$ 
versus $h$ for MSSM parameters resembling the standard point SPS1a', 
random values of the singlet parameters and with the condition of 
${\cal S}_1$ being the LSP. The dominant decay mode for the ${\tilde B}$ 
in all points is ${\tilde B} \rightarrow {\cal S}_1 + \eslash$, with the 
missing energy due to either $J$ or 2$J$ emission. For low values of 
$h$ one can have visible decays of the ${\tilde B}$ reaching $(20-30) \%$, 
for $h$ larger than, for say, $h=0.05$ ${\tilde B}$ decays to 
${\cal S}_1$ plus missing energy with nearly $100 \%$.}
\label{fig:chi3vh}
\end{figure}

If the ${\cal S}$ is the LSP, a bino NLSP decays dominantly to the singlino 
plus missing energy, as is shown in fig. (\ref{fig:chi3vh}). The final 
state can be either ${\cal S}_1 + J$ or ${\cal S}_1 + 2 J$, the latter 
due to the chain ${\tilde B} \to {\cal S}_2 + J \to {\cal S}_1 + 2J$, 
where the 2nd step has always a branching fraction very close to $100 \%$. 
However, a special opportunity arises if $h$ is low. In this case 
$\sum_i Br(\chi^0_3 \simeq {\tilde B} \rightarrow  W^{\pm} + l_i^{\mp})$ 
can easily reach several percent and it becomes possible to test the 
model with the bino decays and the singlino decays {\em at the same time}. 
This would allow a much more detailed study of the model parameters 
than for the more ``standard'' case where only either singlino or 
bino decay visibly. We note that for any fixed value of $h$, 
$\sum_i Br(\chi^0_3 \simeq {\tilde B} \rightarrow  W^{\pm} + l_i^{\mp})$ 
depends mostly on $v_R$ (and to some extend on $v_{\Phi}$). Low values 
if $v_R$ lead to low 
$\sum_i Br(\chi^0_3 \simeq {\tilde B} \rightarrow  W^{\pm} + l_i^{\mp})$ 
as we will discuss next.

Fig. (\ref{fig:Inv_vr}) shows the sum over all at least partially 
visible decay modes of the lightest neutralino versus $v_R$ in GeV, 
for a set of $v_{\Phi}$ values $v_{\Phi} = 10$--40 TeV for the mSUGRA 
parameter point ($m_0=280$~GeV, $m_{1/2}=250$~GeV, $\tan\beta=10$, 
$A_0=-500$~GeV and sgn$(\mu)=+$). This point was constructed to produce 
formally a $\Omega_{\chi^0_1}h^2 \simeq 1$ in case of conserved R-parity, much 
larger than the observed relic DM density \cite{Spergel:2006hy}. The 
left plot shows the case $\chi_1^0 \simeq {\tilde B}$, the right plot 
$\chi_1^0 \simeq {\cal S}$. For ${\tilde B}$, Br(${\tilde B} \to J + \nu$) 
very close to 100 \% are found for low values of $v_R$. This feature 
is independent of the mSugra parameters, see the correspoding figure 
in \cite{Hirsch:2006di}. In this case large statistics becomes necessary 
to find the rare visible neutralino decays, which prove that R-parity is 
broken. The inconsistency between the calculated $\Omega_{\chi^0_1}h^2$ 
and the measured $\Omega_{CDM}h^2$ might give a first indication for 
a non-standard SUSY model.

Figure (\ref{fig:Inv_vr}) to the right shows that the case $\chi_1^0 \simeq 
{\cal S}$ has a very different dependence on $v_R$. We have checked that 
this feature is independent of the mSugra point. For other choices of mSugra 
parameters larger branching ratios for Br(${\cal S} \to J + \nu$) can be 
obtained, but contrary to the bino LSP case, the sum over the invisible 
decay branching ratios never approaches 100 \%.

\begin{figure}[htbp]
\begin{center}
\vspace{5mm}
\includegraphics[width=80mm,height=60mm]{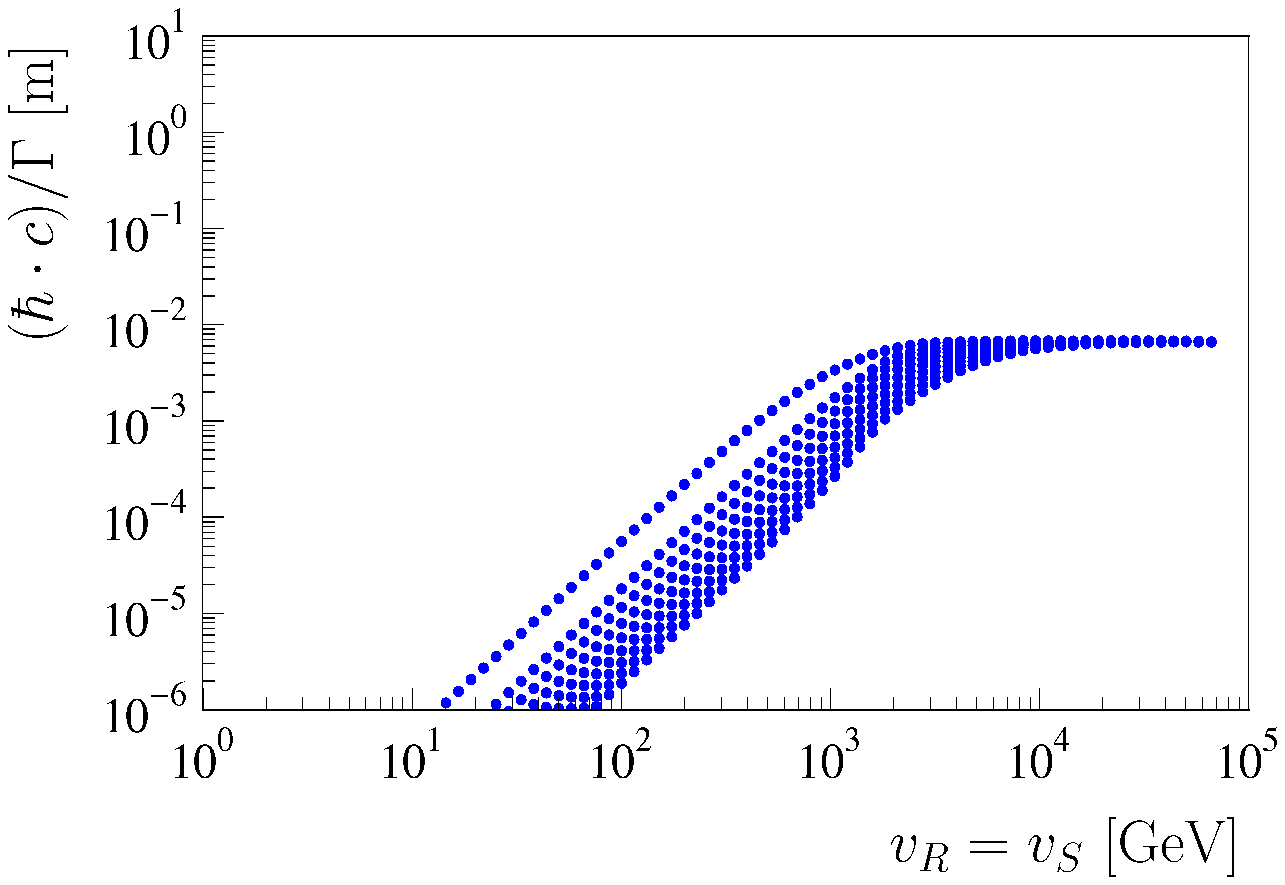}
\includegraphics[width=80mm,height=60mm]{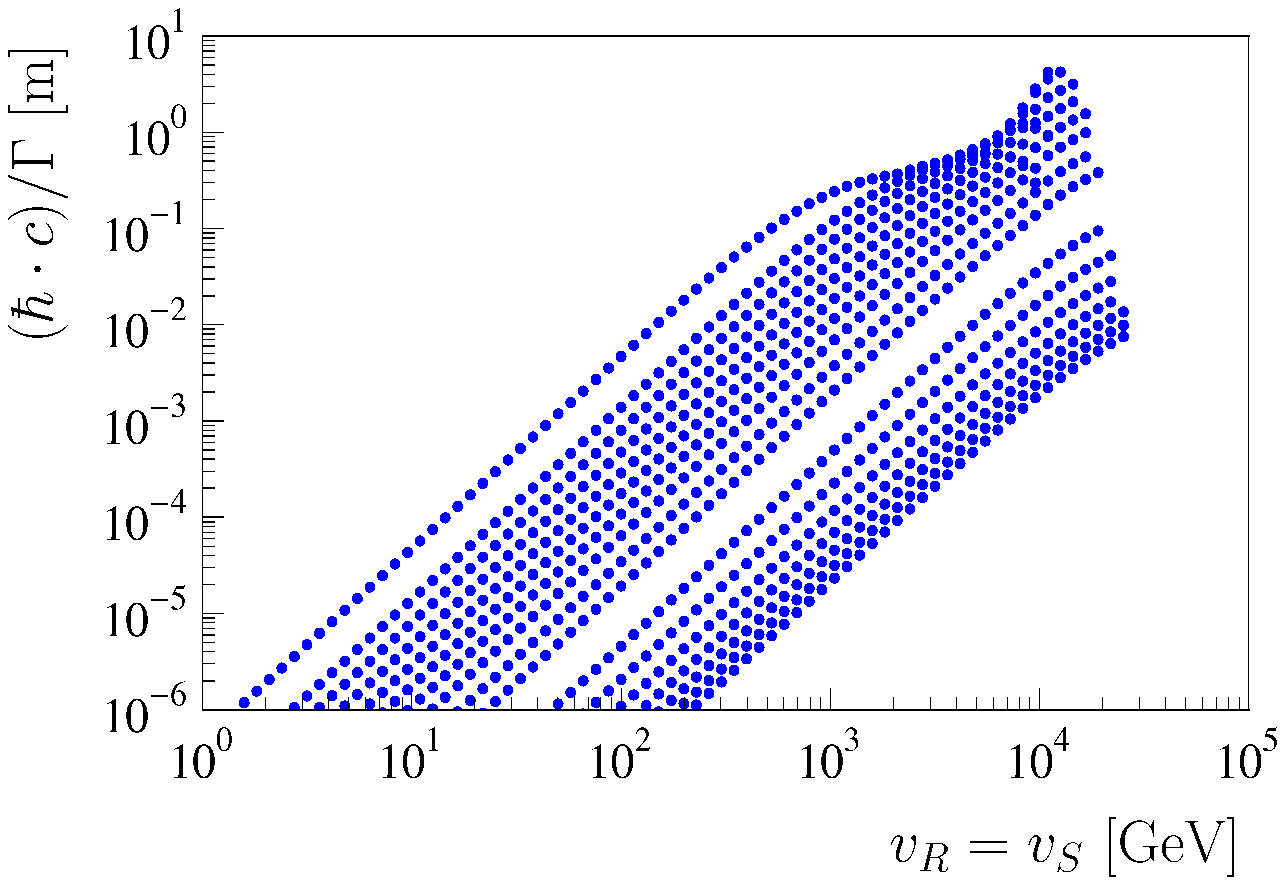}
\end{center}

\caption{Decay length of the lightest neutralino in meter 
versus $v_R$. To the left: Bino LSP; to the right: Singlino 
LSP. All parameters have been chosen as in fig. (\ref{fig:Inv_vr}).}
\label{fig:L_vr}
\end{figure}

Figure (\ref{fig:L_vr}) shows the calculated decay lengths for 
the lightest neutralino for the same choice of parameters as 
shown in fig. (\ref{fig:Inv_vr}). To the left the case $\chi_1^0 \simeq 
{\tilde B}$, to the right $\chi_1^0 \simeq {\cal S}$. Decay 
lengths depend strongly on $v_R$. Singlinos tend to have larger 
decay lengths than binos for the same choice of parameters. 
However, a measurement of the decay length alone is not sufficient 
to decide whether the LSP is a singlino or a bino. If the nature 
of the LSP is known, observing a finite decay length allows a rough 
estimate of the scale $v_R$, or at least to establish a rough lower 
limit on $v_R$.

Summarizing this discussion, it can be claimed that observing a decay 
branching ratio of the LSP into completely invisible final states larger 
than Br$({\chi_1^0} \to \sum [{\rm invisible}])\ge 0.1$ is an indication 
for s-\rpv. Finding Br$({\chi_1^0} \to \sum [{\rm invisible}]) \simeq 100 \%$ 
shows furthermore that the ${\chi_1^0}$ must be a bino and measuring 
Br$({\chi_1^0} \to \sum [{\rm visible}])$ for a bino LSP gives 
an order-of-magnitude estimate of $v_R$.

\subsection{Possible observables to distinguish between Singlino LSP and
bino LSP}

Since bino and singlino LSP decays have, in principle, the same final 
states, simply observing some visible decay products of the LSP does not 
allow to decide the nature of the LSP. In this subsection we will 
schematically discuss some possible measurements, which would allow to 
check for the LSP nature.

As shown above, if the singlino is the LSP and the bino the NLSP, 
one can have that for the bino decays to standard model particles 
compete with the decay to the singlino LSP. If both particles, the 
bino and the singlino LSP have visible decay modes, it is guaranteed 
that the singlino is the LSP. If the bino decays only invisibly to 
the singlino, a different strategy is called for. We discuss two 
examples in the following.

In the following discussion we will replace the neutralino mass
eigenstates by the particles which correspond to their main content
to avoid confusion with indices. At the LHC one will mainly produce
squarks and gluinos which will decay in general in cascades. A typical
example is $\tilde q_L \to q \tilde W$ and the wino decays further to a 
 bino LSP as for example:
\begin{eqnarray}
 \tilde W \to e^+  \tilde e^-
                 \to e^- e^+ \tilde B
                 \to e^- e^+ \mu q \bar{q} \\
\label{eq:bino1}
 \tilde W \to e^+  \tilde e^-
                 \to e^- e^+ \tilde B
                 \to e^- e^+  J \nu
\label{eq:bino2}
\end{eqnarray}
In this case one can measure in principle the neutralino mass from the first
decay chain. In the invariant momentum spectrum of the $e^+ e^-$ pair
the edge must correspond to this mass. 
In the case of the singlino
\begin{eqnarray}
 \tilde W \to e^+  \tilde e^-
                 \to e^- e^+ \tilde B
                 \to e^- e^+  J {\cal S}
                 \to e^- e^+ J \mu q \bar{q} \\
\label{eq:singlino1}
 \tilde W \to e^+  \tilde e^-
                 \to e^- e^+ \tilde B
                 \to e^- e^+  J {\cal S}
                 \to e^- e^+  J J \nu 
\label{eq:singlino2}
\end{eqnarray}
In this case one has on average more missing energy than for a bino LSP. 
However, in both cases one can study spectra combining the jet 
stemming from the squark and the $e^+ e^-$ pair and obtain
information on the masses from the so-called edge variables 
\cite{Allanach:2000kt}. In addition one can use additional variables 
like, for example, $m_{T2}$ \cite{Barr:2003rg,Barr:2007hy,Cho:2007dh}
to obtain information on the LSP mass. Note, that this variable works
also if there are additional massless particles involved, although at 
the expense of available statistics \cite{Barr:private}.  In addition
one can obtain the invariant mass of the LSP from the final state 
$\mu q \bar{q}$. In the case where the LSP has a decay length measurable 
at the LHC, one can separate the latter decay products from the other 
particles in the event and, thus, reduce considerably the combinatorial 
problems associated with the correct assignment of the jets. In the case 
of a bino LSP one would find that all the three different measurements 
yield the same mass for the LSP. In the case of a singlino LSP, on the 
other hand, one would obtain that the LSP mass reconstructed from the 
edge variables does not coincide with the mass reconstructed from the 
$\mu q \bar{q}$ spectrum. This would indicate that there are two different
particles involved. (Such a difference might also be visible in the $m_{T2}$ 
variable.) However, in all cases detailed Monte Carlo studies will be 
necessary to work out the required statistics, etc.

Distinguishing bino and singlino LSPs  will become considerably easier 
at a future international linear collider. In $e^+ e^-$ one can directly 
produce a bino LSP but not a singlino LSP and, thus, the identification of
the correct scenario should be fairly straightforward.

\section{Correlations between LSP decays and neutrino mixing angles}
\label{sec:corr}

Correlations between LSP decays and neutrino mixing angles depend 
on the nature of the LSP. Above we have discussed some possible 
measurements which, at least in principle, allow to distinguish 
bino from singlino LSPs. In this subsection we assume that the 
nature of the LSP is known. 

\subsubsection{Bino LSP}

We note that the following discussion is valid also if the bino is the 
NLSP which, as discussed above, decays with some final, but probably 
small percentage to visible final states. 

\begin{figure}[htbp]
\begin{center}
\vspace{5mm}
\includegraphics[width=80mm,height=60mm]{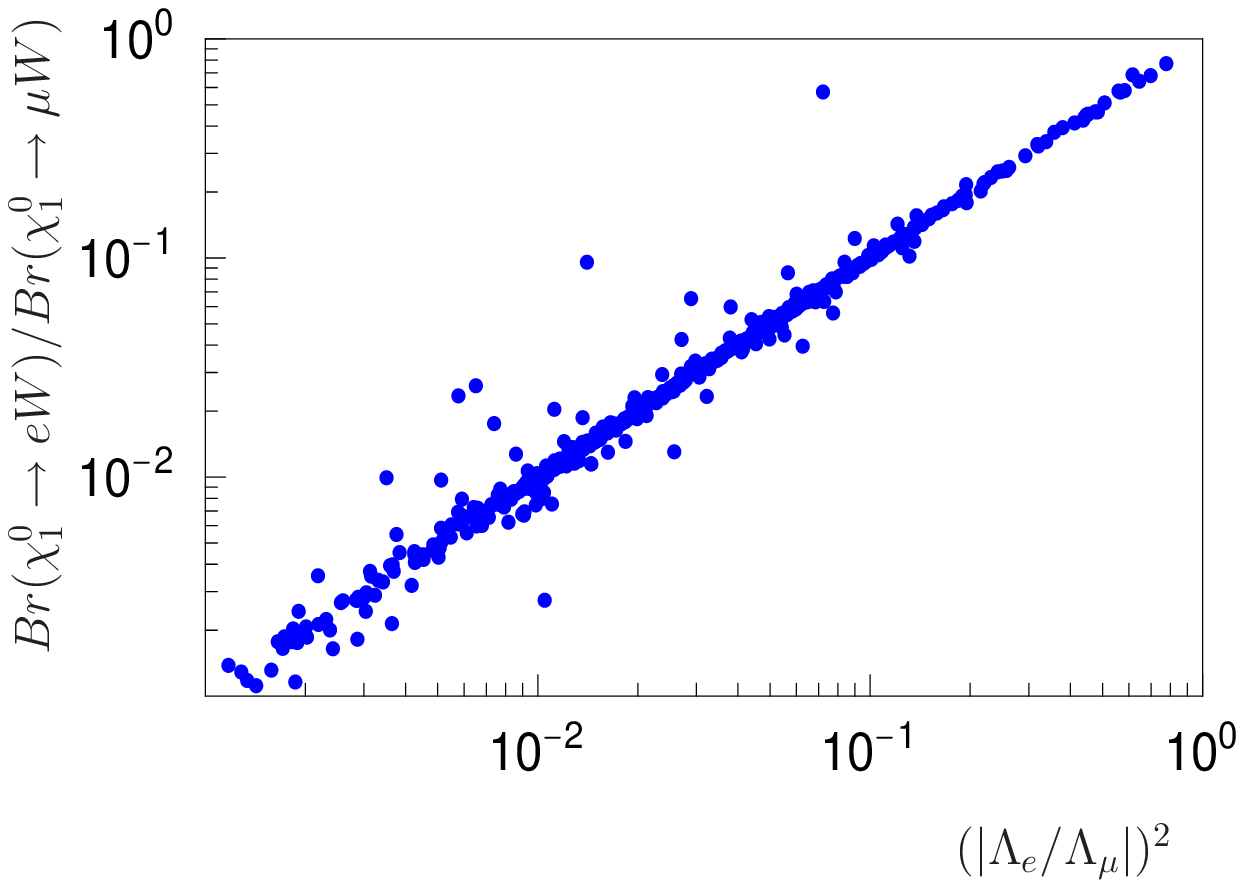}
\includegraphics[width=80mm,height=60mm]{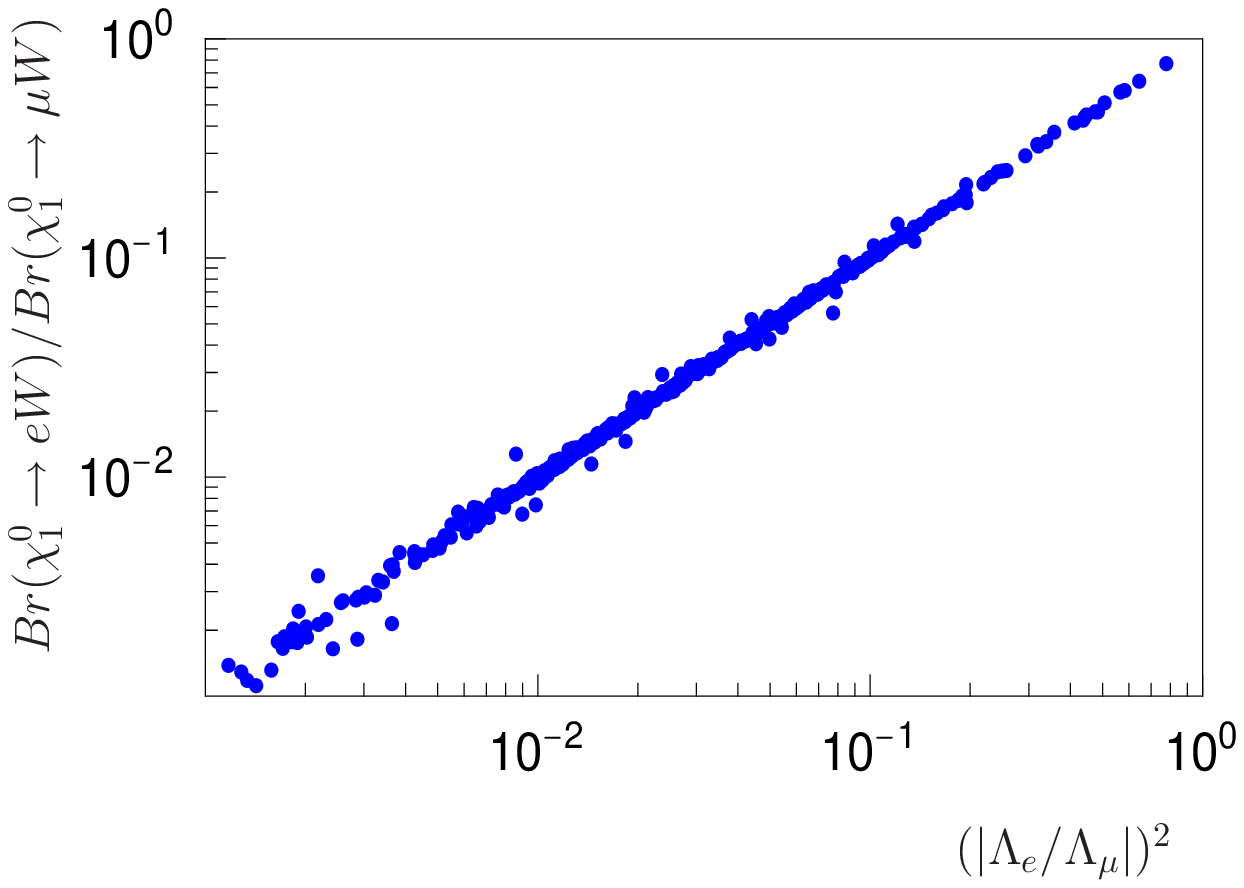}

\end{center}
\vspace{0mm}
\caption{Ratio $\frac{Br(\chi^0_1\rightarrow e W)}
{Br(\chi^0_1\rightarrow \mu W)}$ versus $(\Lambda_{e}/\Lambda_{\mu})^2$ 
for a bino LSP. To the left: ``Bino-purity'' $N_{11}^2 > 0.5$, to the 
right: $N_{11}^2 > 0.9$. All points with $m_{LSP} > m_{W}$.}
\label{fig:BinoeWmuW}
\end{figure}

\begin{figure}[htbp]
\begin{center}
\vspace{5mm}
\includegraphics[width=80mm,height=60mm]{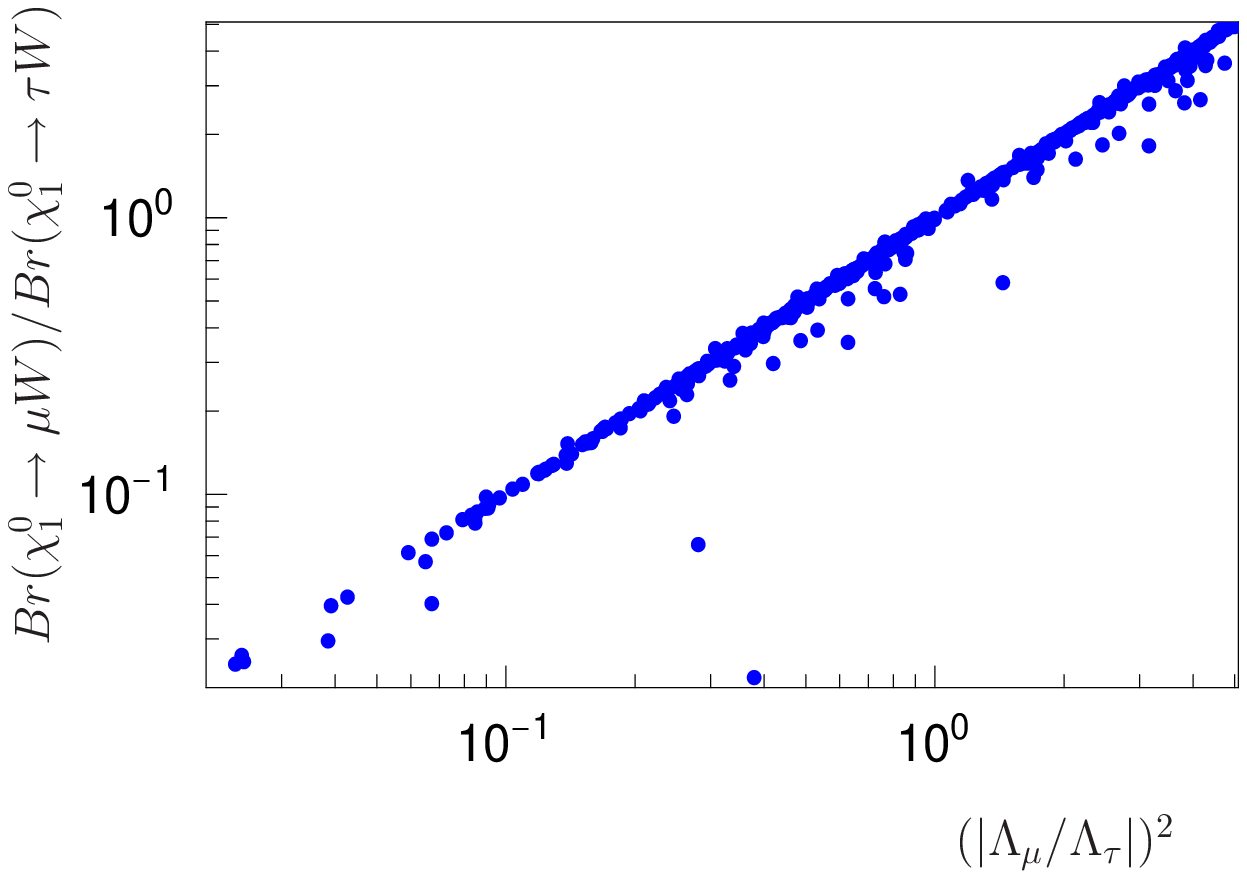}
\includegraphics[width=80mm,height=60mm]{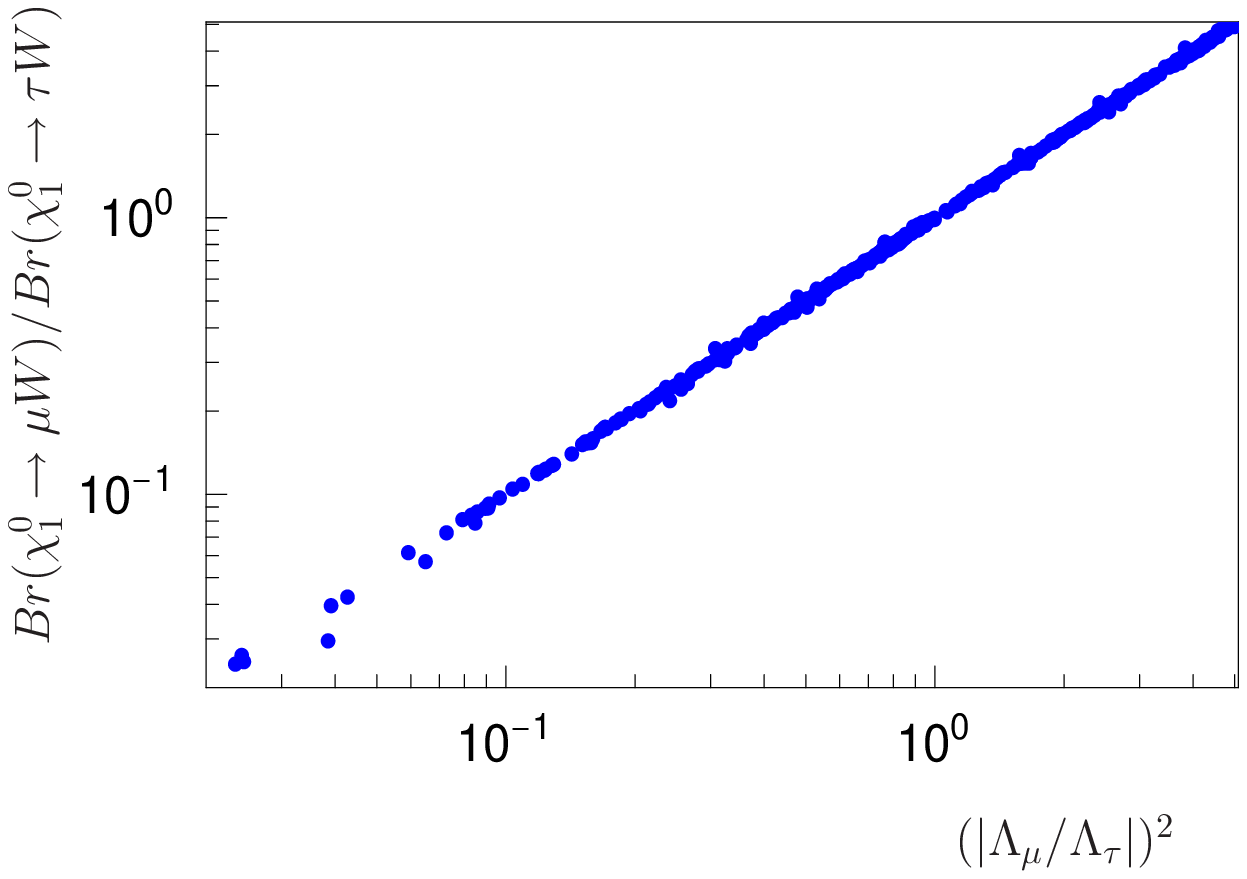}

\end{center}
\vspace{0mm}
\caption{Ratio $\frac{Br(\chi^0_1\rightarrow \mu W)}
{Br(\chi^0_1\rightarrow \tau W)}$ versus $(\Lambda_{\mu}/\Lambda_{\tau})^2$ 
for a bino LSP. To the left: ``Bino-purity'' $N_{11}^2 > 0.5$, to the 
right: $N_{11}^2 > 0.9$.}
\label{fig:BinomuWtauW}
\end{figure}

In explicit bilinear R-parity violation the coupling of the bino component 
of the neutralino to gauge bosons and leptons is completely dominated by 
terms proportional to $\Lambda_i$, as has been shown in \cite{Porod:2000hv}. 
Although the coefficients for the spontaneous model are more complicated, 
see the discussion in section \ref{sec:cpl}, generation dependence for the 
coefficients for the coupling $\chi_1^0-W-l_i$ appear only in the 
terms $\Lambda_i$ and $\epsilon_i$, i.e. $K_{\Lambda}^i$ and  $K_{\epsilon}^i$ 
are independent of the lepton generation. Numerically one finds than 
that the terms proportional to $\Lambda_i$ dominate the $\chi_1^0-W-l_i$ 
coupling for a bino LSP always. This is demonstrated in figs. 
(\ref{fig:BinoeWmuW}) and (\ref{fig:BinomuWtauW}). Here we have 
numerically scanned the mSugra parameter space, with random singlet 
parameters and the additional condition that the LSP is a bino.
For the left (right) figures we have numerically applied the cut 
$N_{11}^2>0.5$ ($N_{11}^2>0.9$). 

Fig. (\ref{fig:BinoeWmuW}) [ (\ref{fig:BinomuWtauW})] shows the ratio 
of branching ratios Br(${\tilde B}\to W + e$)/Br(${\tilde B}\to W + \mu$) 
[Br(${\tilde B}\to W + \mu$)/Br(${\tilde B}\to W + \tau$)] versus 
$(\Lambda_e/\Lambda_{\mu})^2$ [$(\Lambda_{\mu}/\Lambda_{\tau})^2$]. 
To establish a correlation between ratios of $\Lambda_i$ and the 
bino decay branching ratios, a bino purity of $N_{11}^2>0.5$ is 
usually sufficient. The figures demonstrate that the correlations 
get sharper with increasing bino purity.

We have checked that for neutralinos with mass lower than $m_W$ one 
can use ratios of the decays ${\tilde B}\rightarrow l_i qq'$ for the 
different $l_i$ in the same way to perform a measurement of $\Lambda_i$ 
ratios. Plots for this parameter region are rather similar to the ones 
shown for the case ${\tilde B}\rightarrow l_i W$, although with a 
somewhat larger dispersion, and we therefore do not repeat them here. 

\begin{figure}[htbp]
\begin{center}
\vspace{5mm}
\includegraphics[width=80mm,height=60mm]{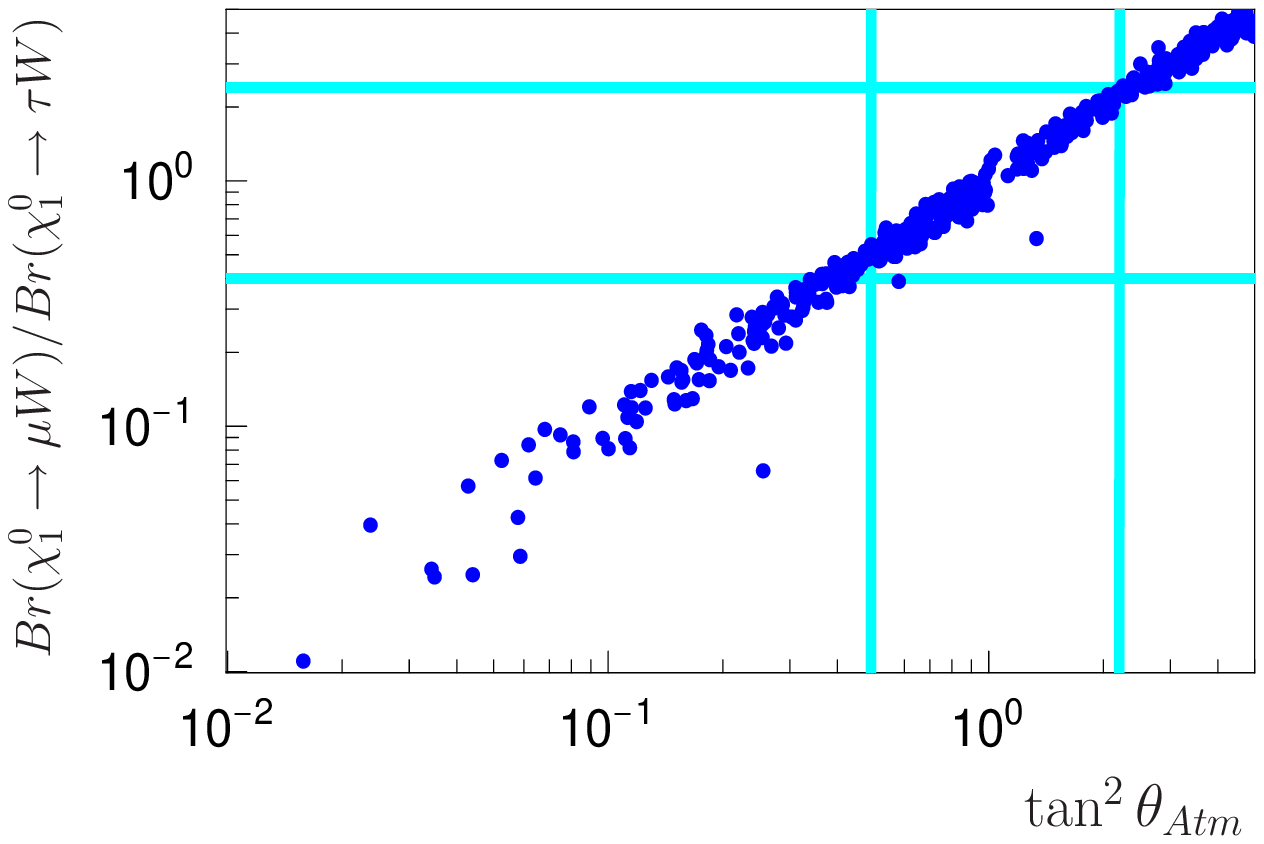}
\includegraphics[width=80mm,height=60mm]{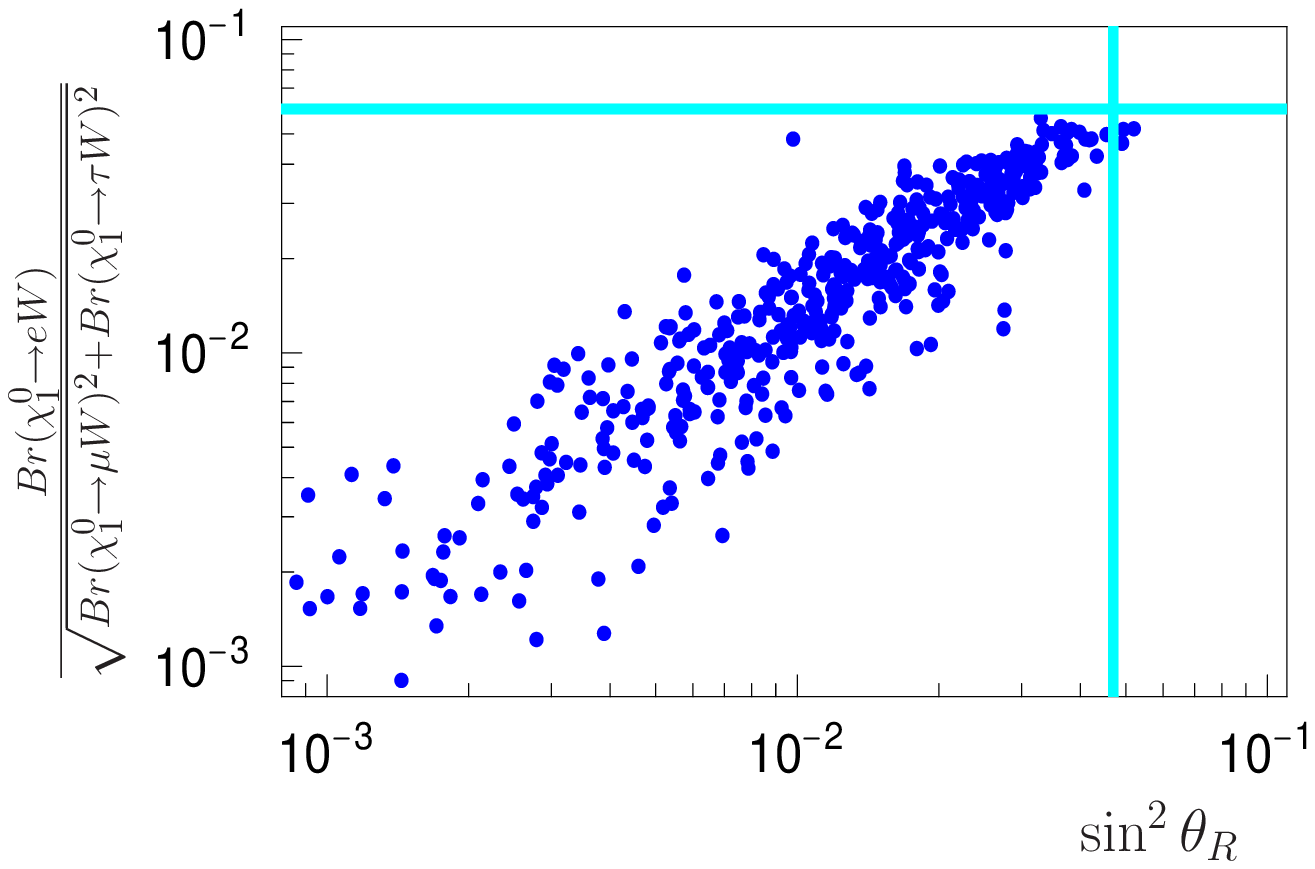}

\end{center}
\vspace{0mm}
\caption{Ratio ${\cal R}_{\mu}= \frac{Br(\chi^0_1\rightarrow \mu W)}
{Br(\chi^0_1\rightarrow \tau W)}$ versus $\tan^2\theta_{Atm}$ (left) 
and ${\cal R}_e= \frac{Br(\chi^0_1\rightarrow e W)}
{\sqrt{Br(\chi^0_1\rightarrow \mu W)^2+Br(\chi^0_1\rightarrow \tau W)^2}}$ 
versus $\sin^2\theta_R$ (right) for a bino LSP. ``Bino-purity'' 
$N_{11}^2 > 0.8$. Vertical lines are the $3 \sigma$ c.l. allowed experimental 
ranges, horizontal lines the resulting predictions for the fit (c1), 
see text.} 
\label{fig:BinoNf1pred}
\end{figure}

With the measurement of ratios of branching ratios different consistency 
checks of the model can be performed. In case (c1), i.e. $\vec\Lambda$
explaining the atmospheric scale, the atmospheric and the reactor angle 
are related to $W+l$ final states, as shown in fig. (\ref{fig:BinoNf1pred}). 
Here we show the ratios ${\cal R}_{\mu} = 
\frac{Br(\chi^0_1\rightarrow \mu W)}{Br(\chi^0_1\rightarrow \tau W)}$ 
versus $\tan^2\theta_{Atm}$ (left) and ${\cal R}_e = 
\frac{Br(\chi^0_1\rightarrow e W)}
{\sqrt{Br(\chi^0_1\rightarrow \mu W)^2+Br(\chi^0_1\rightarrow \tau W)^2}}$ 
versus $\sin^2\theta_R$ (right) for a bino LSP, for an assumed 
bino-purity of $N_{11}^2 > 0.8$. The vertical lines are the 
$3 \sigma$ c.l. allowed experimental ranges (upper bound), horizontal 
lines the resulting predictions for the two different observables 
${\cal R}$. Given the current experimental data, one expects 
$\frac{Br(\chi^0_1\rightarrow \mu W)}{Br(\chi^0_1\rightarrow \tau W)}$ 
in the range $[0.4,2.1]$ and $\frac{Br(\chi^0_1\rightarrow e W)}
{\sqrt{Br(\chi^0_1\rightarrow \mu W)^2+Br(\chi^0_1\rightarrow \tau W)^2}} 
\le 0.06$.

\begin{figure}[htbp]
\begin{center}
\vspace{5mm}
\includegraphics[width=80mm,height=60mm]{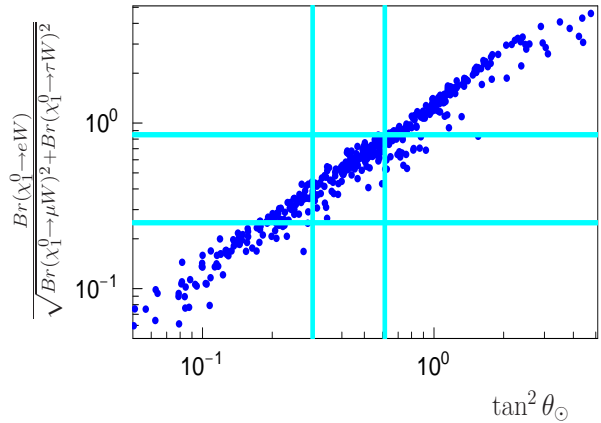}
\end{center}
\vspace{0mm}
\caption{Ratio 
${\cal R}_e= \frac{Br(\chi^0_1\rightarrow e W)}
{\sqrt{Br(\chi^0_1\rightarrow \mu W)^2+Br(\chi^0_1\rightarrow \tau W)^2}}$ 
versus $\tan^2\theta_{\odot}$ for a bino LSP. ``Bino-purity'' 
$N_{11}^2 > 0.8$. Vertical lines are the $3 \sigma$ c.l. allowed experimental 
ranges, horizontal lines the resulting predictions for the fit 
(c2), see text.}
\label{fig:BinoNf3pred}
\end{figure}

Different from fig. (\ref{fig:BinoNf1pred}), in case of (c2), i.e. 
$\vec\Lambda$ explaining the solar scale, the ratio 
$\frac{Br(\chi^0_1\rightarrow e W)}
{\sqrt{Br(\chi^0_1\rightarrow \mu W)^2+Br(\chi^0_1\rightarrow \tau W)^2}}$ 
correlates with $\tan^2\theta_{\odot}$, as shown in fig. 
(\ref{fig:BinoNf3pred}). Here, from the $3 \sigma$ c.l. allowed range 
of the solar angle one expects to find $\frac{Br(\chi^0_1\rightarrow e W)}
{\sqrt{Br(\chi^0_1\rightarrow \mu W)^2+Br(\chi^0_1\rightarrow \tau W)^2}} 
\simeq [0.25,0.85]$. Finding this ratio experimentally to be larger than 
the one indicated by the solar data, i.e. 
$\frac{Br(\chi^0_1\rightarrow e W)}{\sqrt{Br(\chi^0_1\rightarrow \mu W)^2
+Br(\chi^0_1\rightarrow \tau W)^2}} >> 1$, rules out the model as the 
origin of the observed neutrino oscillation data. Similarly a low (high) 
experimental value for this ratio indicates (for a bino LSP) that 
case (c1) [(c2)] is the correct explanation for the two observed 
neutrino mass scales.

\subsubsection{Singlino LSP}

\begin{figure}[htbp]
\begin{center}
\vspace{5mm}
\includegraphics[width=80mm,height=60mm]{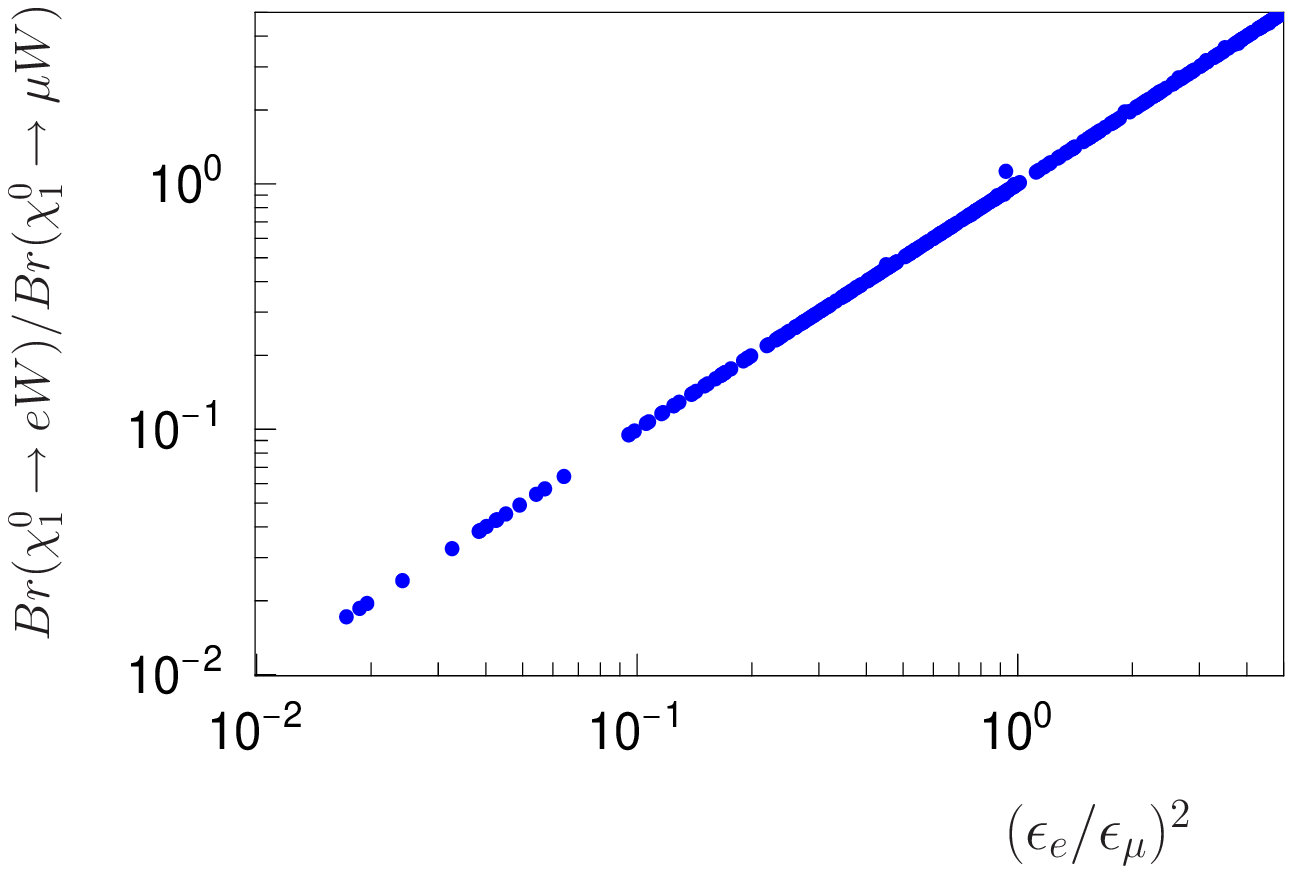}
\includegraphics[width=80mm,height=60mm]{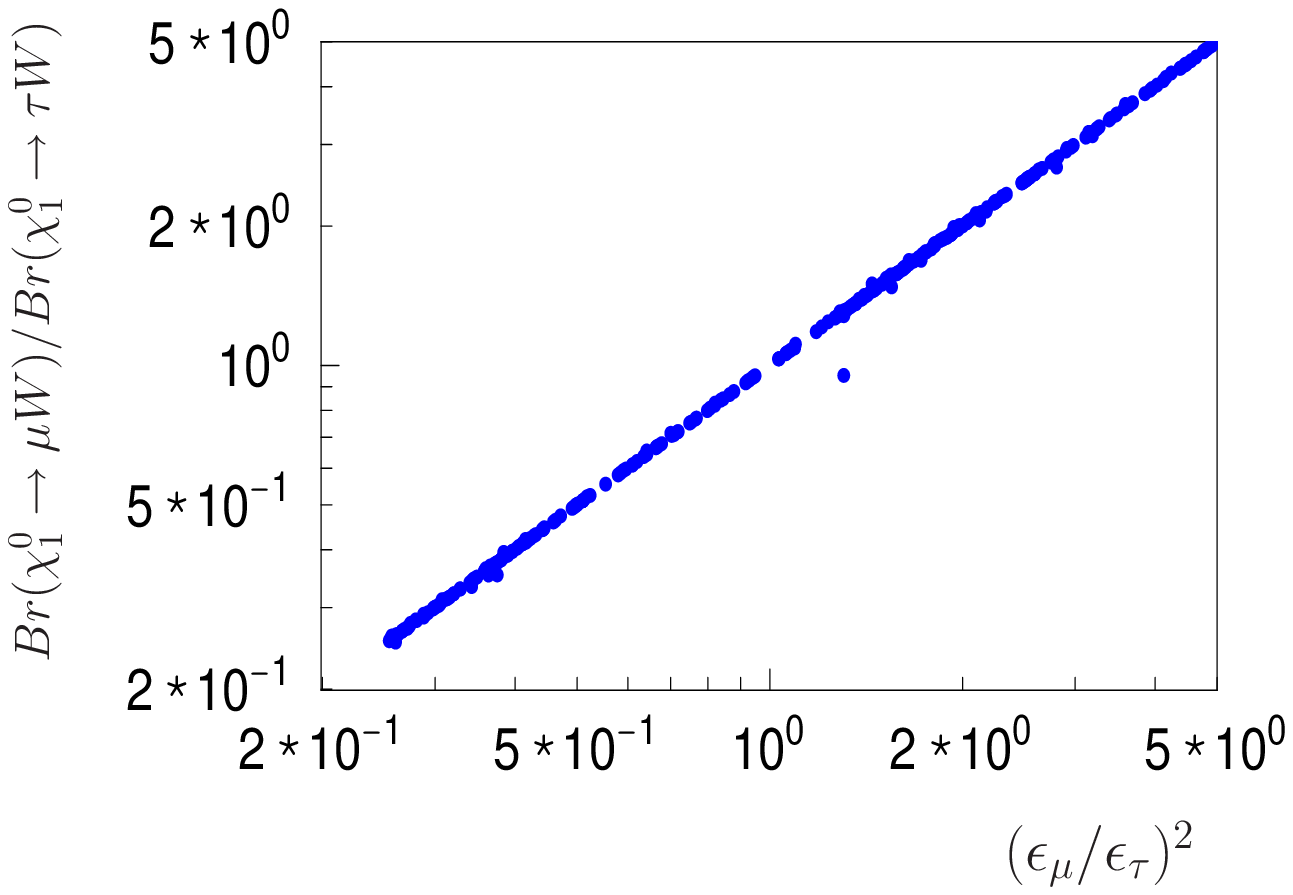}

\end{center}
\vspace{0mm}
\caption{Ratio $\frac{Br(\chi^0_1\rightarrow e W)}
{Br(\chi^0_1\rightarrow \mu W)}$ (left) versus 
$(\epsilon_{e}/\epsilon_{\mu})^2$ and 
$\frac{Br(\chi^0_1\rightarrow \mu W)}
{Br(\chi^0_1\rightarrow \tau W)}$ (right) versus 
$(\epsilon_{\mu}/\epsilon_{\tau})^2$ 
for a ``singlino'' LSP.} 
\label{fig:SngleWmuW}
\end{figure}

Different from the bino LSP case, for singlinos coupling to a lepton 
$l_i$-W pair terms proportional to $\epsilon_i$ dominate by far. 
This is demonstrated in fig. (\ref{fig:SngleWmuW}), where we show the 
ratios $\frac{Br(\chi^0_1\rightarrow e W)}{Br(\chi^0_1\rightarrow \mu W)}$ 
(left) versus $(\epsilon_{e}/\epsilon_{\mu})^2$ and 
$\frac{Br(\chi^0_1\rightarrow \mu W)}{Br(\chi^0_1\rightarrow \tau W)}$ 
(right) versus $(\epsilon_{\mu}/\epsilon_{\tau})^2$ 
for a singlino LSP. Note that mixing between singlinos and 
the doublet neutralinos of the model is always very small, 
unless the singlino is highly degenerate with the bino. 
Consequently singlinos are usually very ``pure'' singlinos and 
the correlations of the $l_i$-W with the $\epsilon_i$ ratios is 
very sharp.

\begin{figure}[htbp]
\begin{center}
\vspace{5mm}
\includegraphics[width=80mm,height=60mm]{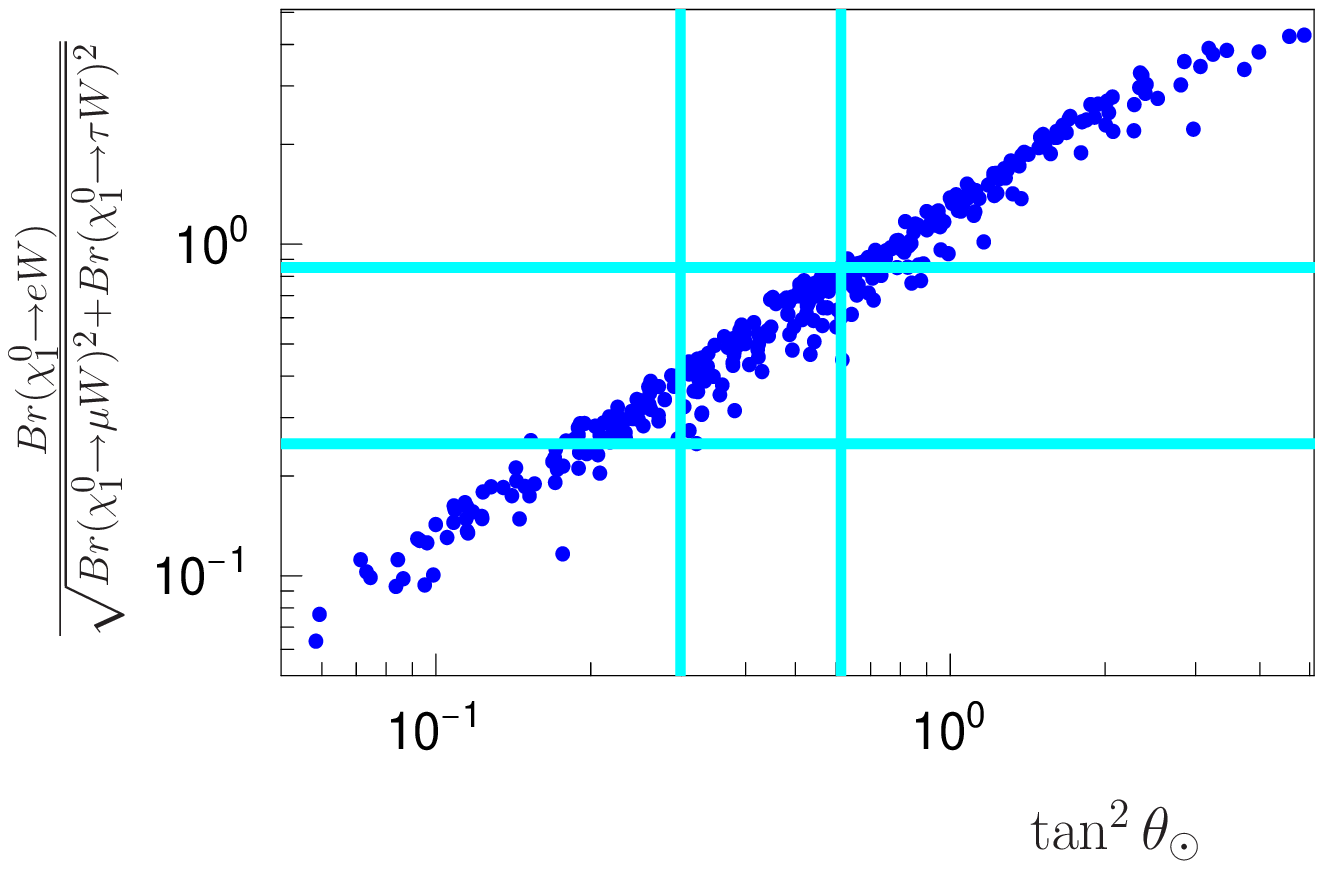}

\end{center}
\vspace{0mm}
\caption{Ratio ${\cal R}_e= \frac{Br(\chi^0_1\rightarrow e W)}
{\sqrt{Br(\chi^0_1\rightarrow \mu W)^2+Br(\chi^0_1\rightarrow \tau W)^2}}$ 
versus $\tan^2\theta_{\odot}$ for a singlino LSP. 
Vertical lines are the $3 \sigma$ c.l. allowed experimental 
ranges, horizontal lines the resulting predictions for the fit (c1), 
see text.}
\label{fig:SnglNf1pred}
\end{figure}

\begin{figure}[htbp]
\begin{center}
\vspace{5mm}
\includegraphics[width=80mm,height=60mm]{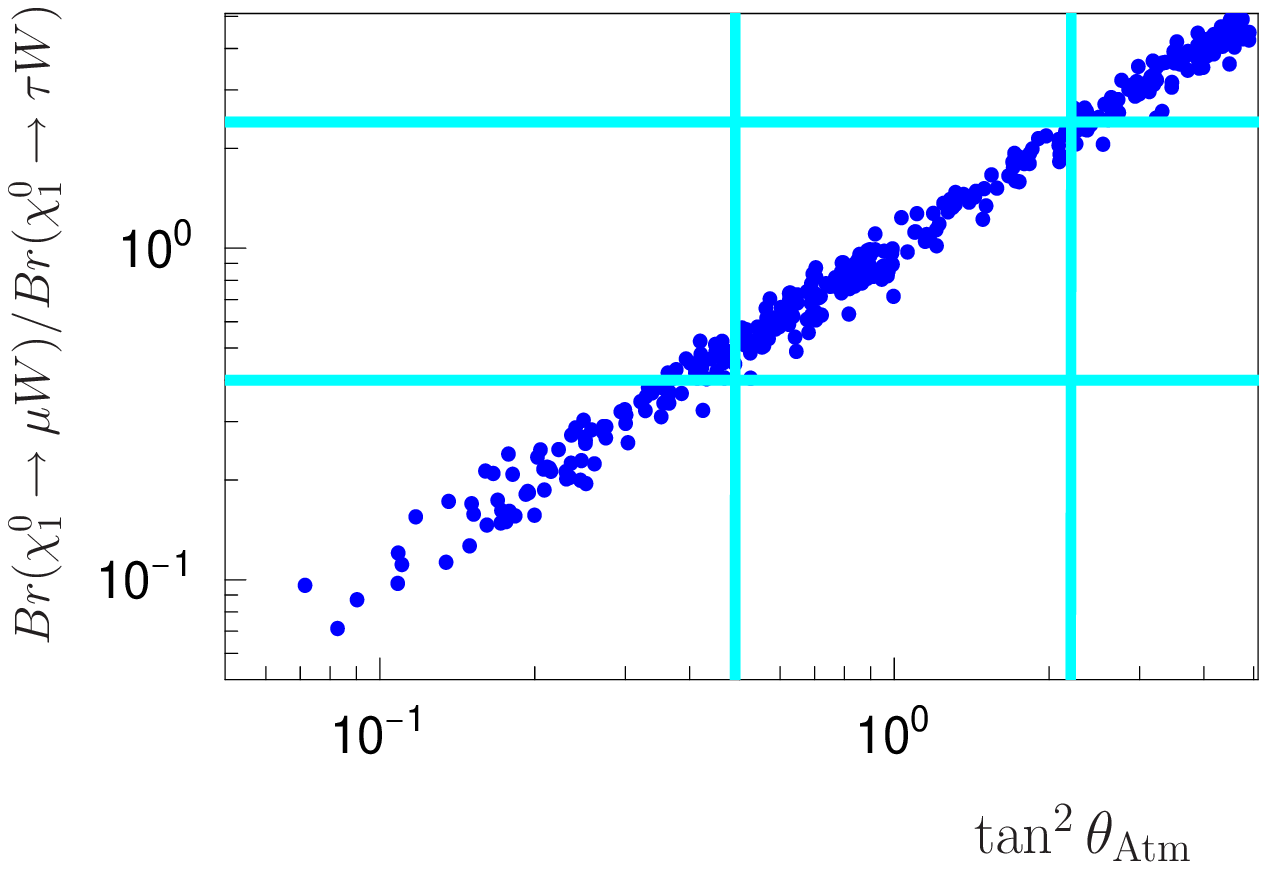}
\includegraphics[width=80mm,height=60mm]{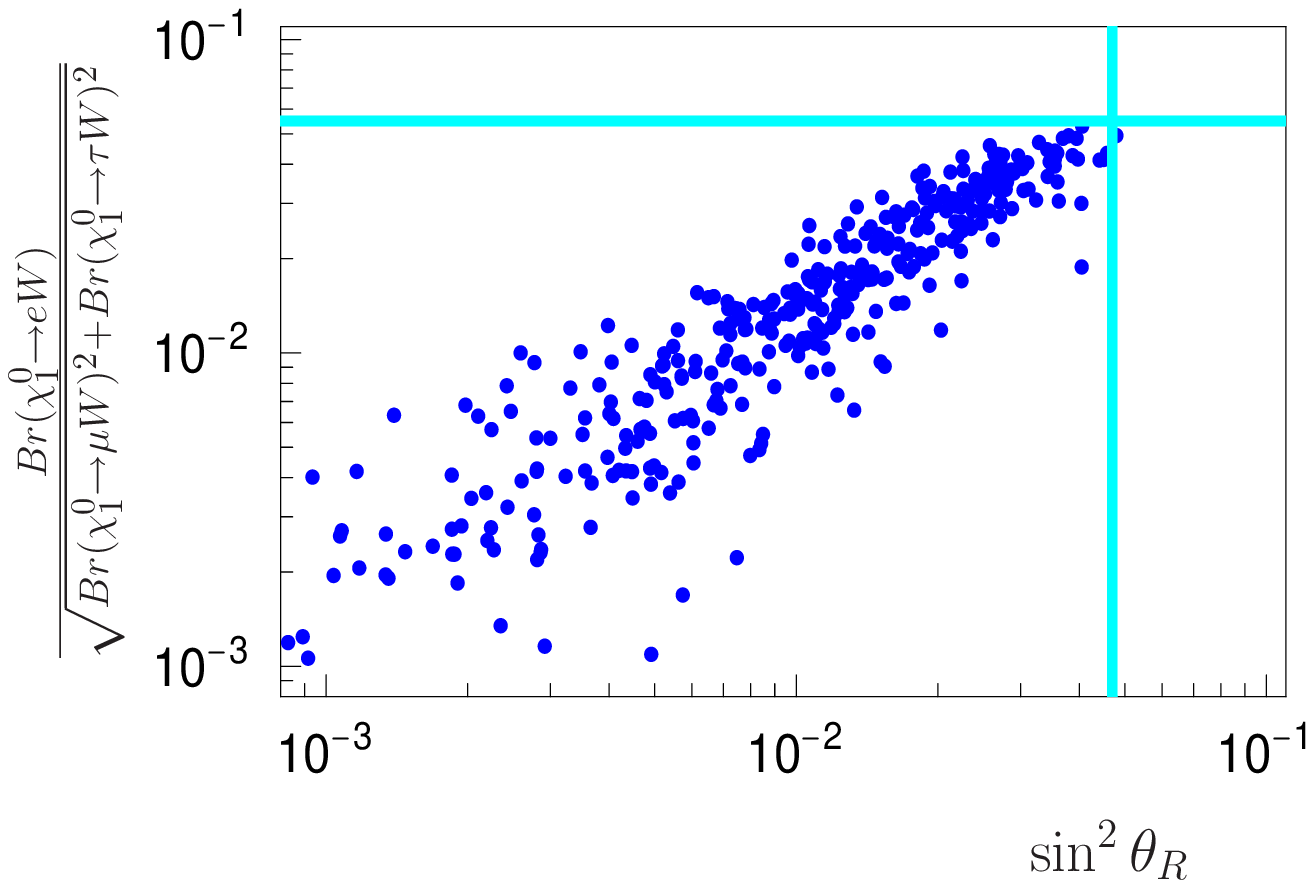}

\end{center}
\vspace{0mm}
\caption{Ratio ${\cal R}_{\mu}= \frac{Br(\chi^0_1\rightarrow \mu W)}
{Br(\chi^0_1\rightarrow \tau W)}$ versus $\tan^2\theta_{Atm}$ (left) 
and ${\cal R}_e= \frac{Br(\chi^0_1\rightarrow e W)}
{\sqrt{Br(\chi^0_1\rightarrow \mu W)^2+Br(\chi^0_1\rightarrow \tau W)^2}}$ 
versus $\sin^2\theta_R$ (right) for a singlino LSP. 
Vertical lines are the $3 \sigma$ c.l. allowed experimental 
ranges, horizontal lines the resulting predictions for the fit (c2), 
see text.}
\label{fig:SnglNf3pred}
\end{figure}

Depending on which case, (c1) or (c2), is chosen to fit the 
neutrino data, the corresponding ratios of branching ratios 
are then either sensitive to the atmospheric and reactor or 
the solar angle. This is demonstrated in figs. (\ref{fig:SnglNf1pred})
and (\ref{fig:SnglNf3pred}). Here, fig. (\ref{fig:SnglNf1pred}) 
shows the correlation of ${\cal R}_e= \frac{Br(\chi^0_1\rightarrow e W)}
{\sqrt{Br(\chi^0_1\rightarrow \mu W)^2+Br(\chi^0_1\rightarrow \tau W)^2}}$ 
with $\tan^2\theta_{\odot}$ for the fit (c1). This result is very 
similar to the one obtained for the fit (c2) and a bino LSP. For 
this reason the nature of the LSP needs to be known, before one 
can decide, whether the measurement of a ratio of branching ratio 
is testing (c1) or (c2).

Figure (\ref{fig:SnglNf3pred}) shows the dependence of 
${\cal R}_{\mu}= \frac{Br(\chi^0_1\rightarrow \mu W)}
{Br(\chi^0_1\rightarrow \tau W)}$ versus $\tan^2\theta_{Atm}$ (left) 
and ${\cal R}_e= \frac{Br(\chi^0_1\rightarrow e W)}
{\sqrt{Br(\chi^0_1\rightarrow \mu W)^2+Br(\chi^0_1\rightarrow \tau W)^2}}$ 
versus $\sin^2\theta_R$ (right) for a singlino LSP, using the 
neutrino fit (c2). Again one observes that this result is very 
similar to the one obtained for a bino LSP and fit (c1). This 
simply reflects that fact, that neutrino angles can be either fitted 
with ratios of $\epsilon_i$ or with ratios of $\Lambda_i$ and 
singlinos couple mostly proportional to $\epsilon_i$, while binos 
are sensitive to $\Lambda_i$. 

In case the singlino is the LSP and the bino, as the NLSP, decays 
with some measurable branching ratios to $W-l_i$, both $\Lambda_i$ 
and $\epsilon_i$ ratios could be reconstructed, which would allow 
for a much more comprehensive test of the model.

\section{Conclusions}
\label{sec:con}
We have studied the phenomenology of a neutralino LSP in a supersymmetric 
model in which neutrino oscillation data is explained by spontaneous 
R-parity violation. We have concentrated the discussion on the case that 
the LSP is either a bino, like in a typical mSugra point, or a singlino 
state, novel to the current model. We have worked out the most important 
phenomenological signals of the model and how it might be 
distinguished from the well-studied case of the MSSM, as well as from 
a model in which the violation of R-parity is explicit. 

There are regions in parameter space, where $\tilde\chi^0$ decays 
invisibly with branching ratios close to 100 \%, despite the smallness 
of neutrino masses. In this limit, spontaneous violation of R-parity 
can resemble the MSSM with conserved R-parity at the LHC and 
experimentalist would have to search for the very rare visible 
decay channels to establish the R-parity indeed is broken.

The perhaps most important test of the model as the origin of the observed 
neutrino masses comes from measurements of ratios of branching ratios 
to $W$-boson and charged lepton final states. Ratios of these decays are 
always related to measured neutrino angles. If SUSY has a spectrum 
light enough to be produced at the LHC, the spontaneous model of 
R-parity violation is therefore potentially testable.

\section*{Acknowledgments}
This work was supported by Spanish grants FPA2005-01269 (MEC), 
ACOMP06/154 (Generalitat Valenciana), by Acciones Integradas 
HA-2007-0090 and by the European Commission Human Potential Program 
RTN network MRTN-CT-2004-503369. A.V. thanks the Generalitat Valenciana 
for financial support. W.P.~is partially supported by the German Ministry 
of Education and Research (BMBF) under contract 05HT6WWA and Fonds zur F\"orderung der
wissenschaftlichen Forschung (FWF) of Austria, project. No. P18959-N16.
 

\bibliographystyle{h-physrev}

\end{document}